\documentclass[11pt]{article}
\usepackage{algorithm}
\usepackage{algorithmic}
\usepackage[T1]{fontenc}
\usepackage{latexsym}
\usepackage{booktabs}
\usepackage{multirow}
\usepackage{cite}
\usepackage{enumitem}
\usepackage{amssymb,mathrsfs}
\usepackage{booktabs}
\usepackage[justification=centering]{caption}
\usepackage{subfigure}
\usepackage{float}
\usepackage{color}
\usepackage{amsmath}
\usepackage[colorlinks,
            linkcolor=blue,
            anchorcolor=red,
            citecolor=green]{hyperref}
\usepackage{bm}
\usepackage{graphicx}
\usepackage{float}
\usepackage{subfigure}
\usepackage{wrapfig}
\usepackage{amsthm}
\usepackage{epstopdf}
\usepackage{multirow}
\usepackage[table,xcdraw]{xcolor}
\usepackage{booktabs}

\allowdisplaybreaks[4]

\newtheorem{theorem}{Theorem}[part]

\renewcommand{\baselinestretch}{1.0}
\begin{document}
\title{Poincar\'{e} Heterogeneous Graph Neural Networks for Sequential Recommendation}


\author{Naicheng Guo\thanks{Ant Group, Beijing, China. Email:\{guonaicheng.gnc, liuxiaolei.lxl, lishaoshuai.lss\}@alibaba-inc.com, \{qiongxu.mqx, hanbing.hanbing, jefflittleguo.gxb\}@antgroup.com. }
\and
Xiaolei Liu$^\ast$
\and 
Shaoshuai Li$^\ast$
\and
Qiongxu Ma$^\ast$
\and
Kaixin Gao \thanks{School of Mathematics, Tianjin University, Tianjin, China.
Email: gaokaixin@tju.edu.cn.}
\and
Bing Han$^\ast$
\and
Lin Zheng\thanks{Department of Computer Science, Shantou University, Shantou, China. Email:lzheng@stu.edu.cn. }
\and
Xiaobo Guo$^\ast$
}

\date{}

\maketitle

\begin{abstract}
 
Sequential recommendation (SR) learns users' preferences by capturing the sequential patterns from users' behaviors evolution. 
As discussed in many works, user-item interactions of SR generally present the intrinsic power-law distribution, which can be ascended to hierarchy-like structures. 
Previous methods usually handle such hierarchical information by making user-item sectionalization empirically under Euclidean space, which may cause distortion of user-item representation in real online scenarios. 
In this paper, we propose a Poincar\'{e}-based heterogeneous graph neural network named PHGR to model the sequential pattern information as well as hierarchical information contained in the data of SR scenarios simultaneously. 
Specifically, for the purpose of explicitly capturing the hierarchical information, we first construct a weighted user-item heterogeneous graph by aliening all the user-item interactions to improve the perception domain of each user from a global view.
Then the output of the global representation would be used to complement the local directed item-item homogeneous graph convolution.
By defining a novel hyperbolic inner product operator, the global and local graph representation learning are directly conducted in Poincar\'{e} ball instead of commonly used projection operation between Poincar\'{e} ball and Euclidean space, which could alleviate the cumulative error issue of general bidirectional translation process. Moreover, for the purpose of explicitly capturing the sequential dependency information, we design two types of temporal attention operations under Poincar\'{e} ball space.
Empirical evaluations on datasets from the public and financial industry show that PHGR outperforms several comparison methods.
\end{abstract}

\maketitle
\section{Introduction}
SR has become a research hotspot in recent years due to its excellent performance in predicting users' preferences by building models from users' historical interaction data. 
The development of SR has made significant progress, from the early Markov-chain-based (MC) methods  \cite{rendle2010factorizing,wang2015learning,he2016fusing}, Recurrent-Neural-Networks-based (RNN) methods \cite{hidasi2015session,li2017neural,cui2018mv}, attention-based mechanism \cite{kang2018self,zheng2020sentiment,fan2021lighter} to Graph-Neural-Networks-based (GNNs) approaches due to their remarkable ability on representing the coherence information contained in the structured interaction data \cite{wu2019session,wu2019personalizing,xu2019graph}. In this paper, we focus on developing and improving GNNs-based approach in SR scenarios.

As discussions in many works \cite{ravasz2003hierarchical,wang2015learning,wang2018exploring,ma2019hierarchical,li2020hierarchical}, 
user-item interactions of SR generally present the intrinsic power-law distribution, which means that a majority of users/items have very few interactions and a few users/items have a huge number of interactions. As a result, the degree coefficient of the graph data derived from user-item interactions also exhibits power-law distribution. Such phenomenon can be approximated described as the `hierarchical' structure according to the degree coefficients slicing. Nodes with similar degree coefficients are at the same level. For example, in the first level, nodes have the highest degree coefficients (core members), while those in the last level have the lowest degree coefficients(marginal members). The number of nodes grows exponentially as the degree of nodes increases. In other words, the number of nodes grows exponentially with the increasing of distance from the hierarchy origin.

Various approaches have been introduced to capture the underlying hierarchical properties under Euclidean space, such as slicing the historical behaviors via long/short time periods or categorizing interacted items empirically. 
However, such Euclidean operations may cause distortion of user-item representation and unspectacular recommendation performance in real online scenarios \cite{bronstein2017geometric,sala2018representation}. 
As Gulcehre \cite{gulcehre2018hyperbolic} suggested, such hierarchical user-item interaction properties cannot be efficiently represented in Euclidean space but are capable of being exploited in hyperbolic space. As demonstrated in the previous papers \cite{krioukov2008efficient,krioukov2010hyperbolic}, the hierarchy-like geometry of power-law node degree distribution and hyperbolic spaces are intimately correlated. As if nodes distribute approximately uniformly in a hidden hyperbolic space, then the number of nodes at distance $R$ from any reference point, e.g., hidden hierarchy origin denoted as the nodes with highest degree coefficients, grows exponentially with $R$. 
In other words, the manifold hyperbolic space has a larger capacity compared to Euclidean space given the same radius and thus more nodes could be contained and effectively represented. More generally, by the definition of the curvature, hyperbolic spaces with negative curvature expand faster than Euclidean spaces with zero curvature. Specifically, while Euclidean spaces expand polynomially, hyperbolic spaces expand exponentially.  In the hyperbolic plane, for example, the length of the circle and the area of the disc of hyperbolic radius $R$ are
\[
l(R) = 2\pi \sinh R,
\]
\[
s(R) = 2\pi(\cosh R-1).
\]

In summary, hierarchy-like geometry of power-law node degree distribution needs an exponential amount of space for expanding, and hyperbolic geometry just has it.

Recently, many researchers have tried to apply hyperbolic learning to recommendation
systems\cite{chamberlain2019scalable,Ky2020,ch2021,li2021hsr}. Motivated by these works, introducing hyperbolic representation along with graph learning to SR would be a promising approach by capturing the coherent and hierarchical information in the data simultaneously. However, this idea still faces two critical technical challenges. 



Firstly, the commonly used graph representation learning deriving from users' historical behaviors becomes more difficult under hyperbolic space as coherent and hierarchical information should be seized simultaneously. 
As one user’s preference may be inferred from similar users, many recent studies \cite{lin2020fissa,wang2020knowledge,wang2020global} have shown the significance of involving clustering users’ preference in GNNs-based SR methods,  which means not only the target user’s historical behaviors are considered but also the related transition items from other users are incorporated.
Such above incorporation should essentially be resolved under hyperbolic space, which brings higher computation complexity.
\begin{figure}[ht] 
\centering 
\includegraphics[width=0.46\linewidth]{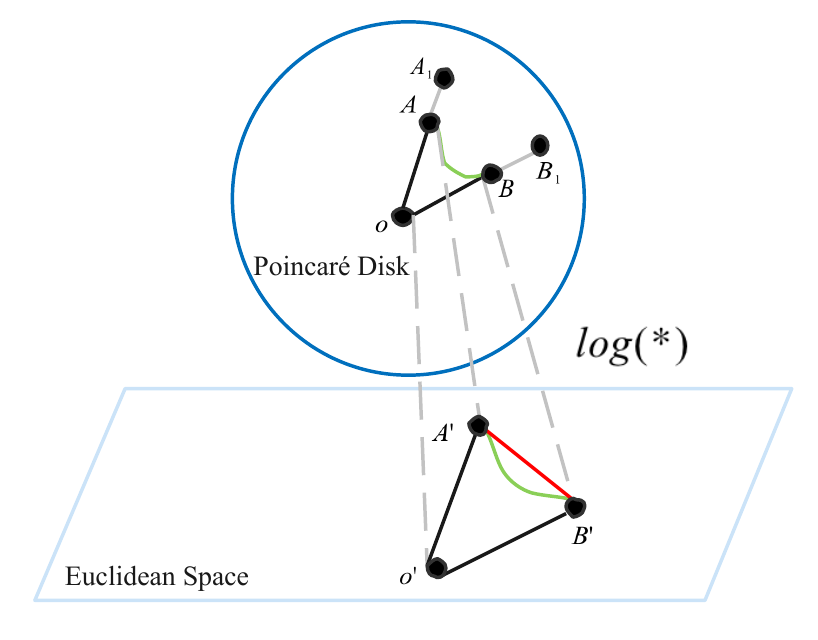}
\caption{The projection process between Euclidean space and hyperbolic space.} 
\label{gap1} 
\end{figure}

Secondly, computing the matching score of one user and corresponding items directly in hyperbolic space is a challenging task as inner product operation under hyperbolic space still lacks a powerful mathematical foundation. The commonly used compromise solution is to carry out such matching computation under Euclidean space by involving two separate stages, which are projection stage and calculation stage. Specifically, hyperbolic vectors are firstly projected into Euclidean space and then the matching calculation based on Euclidean inner product \cite{kang2018self, yi2019sampling} is conducted as follows:

\begin{equation}\label{inner_product}
    \langle\mathbf{x},\mathbf{y}\rangle =\frac{1}{2}(
\|\mathbf{x}\|^2+\|\mathbf{y}\|^2-\|\mathbf{x}-\mathbf{y}\|^2
).
\end{equation}
However, the above two-stage approach would cause calculation error as distance $\|\mathbf{x}-\mathbf{y}\|^2$ in Eq. (\ref{inner_product}) under Euclidean space deviates from the corresponding intrinsic geodesic distance under hyperbolic space. An intuitional example is shown in Figure \ref{gap1}, where $OA$ and $OB$ denote the embedding vectors in Poincar\'{e} ball and $O'A'$, $O'B'$ denote the corresponding projected vectors in Euclidean space. 
The inner product result of $O'A'$ and $O'B'$ based on Eq. (\ref{inner_product}) would be used to represent the hyperbolic inner product result of $OA$ and $OB$ approximatively.
However, such operation is inexact as $\|\mathbf{OA}\|$ $=$ $\|\mathbf{O'A'}\|$ and $\|\mathbf{OB}\|$ $=$ $\|\mathbf{O'B'}\|$, while $\|\mathbf{A'B'}\|$ $\neq$ $\|\mathbf{AB}\|$. $\|\mathbf{A'B'}\|$ and $\|\mathbf{AB}\|$ represent Euclidean distance and hyperbolic geodesic distance, respectively. In other words, 
this two-stage operation cannot satisfy the geodesic distance relationship of two embedding vectors in hyperbolic space. To solve this issue, Ivana et al. \cite{balazevic2019multi} introduced a new matching score function which absorbs squared norms $\|\mathbf{x}\|$ and $\|\mathbf{y}\|$ into biases $\mathbf{b_x}$ and $\mathbf{b_y}$ and replace the Euclidean distance $\|\mathbf{x}-\mathbf{y}\|$ with the Poincar\'{e} distance $d_{\mathbb{B}^n}^2(\mathbf{x},\mathbf{y})$. And the biases $\mathbf{b_x}$ and $\mathbf{b_y}$ determine the radius of a hypersphere decision boundary centered at the embedding of subject. Since biases are subject and object entity-specific, each subject-object pair induces a different decision boundary. But such method would increase training parameters significantly and can't be a substitute for the inner product. But such method would increase training parameters significantly. Therefore, how to design a novel inner product operation under hyperbolic space without involving additional parameters becomes another practical challenge.

To address the aforementioned challenges, we propose a novel Poincar\'{e} Heterogeneous Graph Neural Networks for Sequential Recommendation, namely PHGR. PHGR is a principled SR framework that employs hyperbolic Poincar\'{e} ball to adequately exploit the coherent and hierarchical information contained in one user's local and clustering global behaviors. Specifically, we first create a global heterogeneous graph based on all users’ preferences within the dataset to improve the perception domain of each user by embedding user-item representation into Poincar\'{e} ball to reserve their hierarchical and coherent properties. Then, the output of the global representation is used to enhance the representation learning of each users’ directed homogeneous local behavior graph in Poincar\'{e} ball. Thirdly, we introduce a hyperbolic graph attention mechanism oriented for the global heterogeneous graph and local heterogeneous graph processes. Moreover, we define a novel hyperbolic inner product operation to reduce the cumulative error of general bidirectional transition process between Poincar\'{e} ball and Euclidean space. Overall, our major contributions can be summarized as follows:
\begin{itemize}
\item We propose a novel framework PHGR, which adequately captures the hierarchical and coherent information contained in one user's local and global behaviors under Poincar\'{e} ball space along with a newly designed hyperbolic graph attention mechanism for SR tasks. To the best of our knowledge, our method is the first one to simultaneously extract hierarchical information from one user's local and global reception field.
We theoretically defined a novel inner product operation under Poincar\'{e} ball to reduce the cumulative error of general bidirectional translation process between Poincar\'{e} ball and Euclidean space.
\item Extensive industrial and public experimental results indicate that our proposed PHGR significantly outperforms the state-of-the-art baselines.
\end{itemize}

\section{Related Work}
In this section, we review existing studies on sequential recommender systems and hyperbolic representation learning, which are two fields related to our study.
In this section, we summarize the related studies on sequential recommender system, which is a field related to our study.
\subsection{Sequential Recommender Systems}
The methods for SR can be roughly categorized into four categories. The original MC-based method exploits one user's next preference based on his/her last behavior \cite{rendle2010factorizing,wang2015learning,he2016fusing}. Then, RNN-based method is proposed to predict one user's interests based on strict temporal behaviors \cite{hidasi2015session,li2017neural,cui2018mv}. On the basis of RNN-based method, attention-based method is introduced to simulate unidirectional message transformation between consecutive elements within a sequence by several novel attention mechanisms \cite{kang2018self,zheng2020sentiment,fan2021lighter}. However, the aforementioned three categories of methods cannot handle the intricate item transition relationships. To address this problem, GNN-based methods are widely used to detect the implicit and explicit coherent information within the complicated user-item interactions. Some studies \cite{wu2019session,xu2019graph,wu2019personalizing} utilize digraphs derived from historical behaviors to capture users' coherent preference patterns. Beyond the above works, graph memory \cite{ma2020memory} and information lossless \cite{chen2020handling} mechanism are introduced to further reduce the loss of information aggregation process.
Aforementioned, the prior existing GNN-based SR methods do not consider long-standing hierarchical information contained in one user's historical behaviors.

\subsection{Poincar\'{e} Embedding in Hyperbolic Space}
Recent investigations have revealed that the structure of most complicated data is highly non-Euclidean \cite{bronstein2017geometric,duin2010non}. Hyperbolic space is a strikingly fascinating concept that sparks a lot of research \cite{sala2018representation,tifrea2018poincar}. The Poincar\'{e} model and the Lorentz model of the hyperbolic space have been widely used in a variety of practical applications due to its high representation ability \cite{nickel2017poincare,ganea2018hyperbolic,chami2019hyperbolic,mathieu2019continuous,ovinnikov2019poincar,skopek2019mixed,chami2020low,choudhary2021self}. In \cite{nickel2017poincare}, the hierarchical representations of symbolic data were developed by embedding symbolic data in Poincar\'{e} ball. Then, the hyperbolic neural networks were proposed in \cite{ganea2018hyperbolic}, which merges the formalism of Möbius gyrovector spaces with the Riemannian geometry of the Poincar\'{e} model of hyperbolic spaces. Afterwards, the Hyperbolic Graph Convolutional Network scheme was presented in \cite{chami2019hyperbolic}. Its core idea is to combine the Lorentz model of the hyperbolic space with graph convolutional networks to learn graph representations. Also, the concept of hyperbolic space has been widely applied to the variational autoencoder \cite{mathieu2019continuous,ovinnikov2019poincar,skopek2019mixed}. In addition, hyperbolic knowledge graph embedding was developed in \cite{chami2020low,choudhary2021self} to capture both logical and hierarchical patterns. The advantages and usefulness of hyperbolic space in learning hierarchical structures of complicated relational data have been demonstrated in these studies.

Many researchers attempt to use hyperbolic learning in recommender systems after noticing the potential of hyperbolic space in learning complicated relationships between item and user \cite{chamberlain2019scalable,feng2020hme,mirvakhabova2020performance,li2021shr,wang2021hypersorec}. In \cite{chamberlain2019scalable}, Chamberlain et al. also pointed that millions of users could benefit from the recommender system based on hyperbolic space. A non-Euclidean embedding model \cite{feng2020hme} was developed and applied to next-POI recommendation. Besides, the Hyperbolic Geometry Model for Top-K recommendation was given in \cite{mirvakhabova2020performance}. Hyperbolic Social Recommender presented by \cite{li2021hsr} uses hyperbolic geometry to improve performance of SR. As discussed in \cite{wang2021hypersorec}, a novel graph neural network framework, namely HyperSoRec, combines hyperbolic learning with social recommendation to solve the social recommendation tasks. Note that, in hyperbolic space, there is no solid mathematical foundation to compute the inner product, the aforementioned hyperbolic learning methods need to map their embedding from the hyperbolic space to the Euclidean space whenever they need to perform the inner production between embedding, which may cause a gap.

Many investigations have revealed that the structure of most complicated data is highly non-Euclidean \cite{bronstein2017geometric,duin2010non}. Hyperbolic space is a fascinating concept that sparks a lot of research \cite{sala2018representation,tifrea2018poincar}. 
The Poincar\'{e} model and the Lorentz model of the hyperbolic space have been used in a variety of applications due to its high representation ability \cite{nickel2017poincare,ganea2018hyperbolic,chami2019hyperbolic,mathieu2019continuous,ovinnikov2019poincar,skopek2019mixed,chami2020low,choudhary2021self}. 
For example, by embedding symbolic data in Poincar\'{e} ball, the literature \cite{nickel2017poincare} developed hierarchical representations of symbolic data.
By merging the formalism of Möbius gyrovector spaces with the Riemannian geometry of the Poincar\'{e} model of hyperbolic spaces, \cite{ganea2018hyperbolic} presented hyperbolic neural networks. 
\cite{chami2019hyperbolic} developed the Hyperbolic Graph Convolutional Networks, which combines Lorentz model of the hyperbolic space with graph convolutional networks to learn graph representations.
\cite{mathieu2019continuous,ovinnikov2019poincar,skopek2019mixed} proposed variational autoencoder based on hyperbolic space.
Hyperbolic knowledge graph embedding \cite{chami2020low,choudhary2021self}, which captures both logical and hierarchical patterns, has also been proposed. 
The advantages and usefulness of hyperbolic space in learning hierarchical structures of complicated relational data have been demonstrated in these studies. 

Many researchers attempt to use hyperbolic learning in recommender systems after noticing the potential of hyperbolic space in learning complicated relationships between item and user \cite{chamberlain2019scalable,feng2020hme,mirvakhabova2020performance,li2021hsr,wang2021hypersorec}. \cite{chamberlain2019scalable} also pointed that millions of users could benefit from the hyperbolic recommender system based on hyperbolic space. \cite{feng2020hme} suggests a non-Euclidean embedding model for the next-POI recommendation. 
\cite{mirvakhabova2020performance} gave a Hyperbolic Geometry Model for Top-K recommendation.
Hyperbolic Social Recommender presented by \cite{li2021hsr} uses hyperbolic geometry to improve performance. 
HyperSoRec is a novel graph neural network framework that combines hyperbolic learning with social recommendation for social recommendation tasks \cite{wang2021hypersorec}.
Due to the lack of inner product operations in hyperbolic space in existing studies, the aforementioned methods need to map their embedding from the hyperbolic space to the Euclidean space whenever they need to perform the inner production between embedding, which may cause a gap.

\section{Preliminaries}
In this section, we first present the notations and problem formulation. Then, we describe two types of graphs related to the present PHGR. Lastly, we introduce basic knowledge about hyperbolic geometry.
\subsection{Notations and Problem Formulation}
In general, given a user behavior sequence, i.e., a chronological item clicking sequence of that user, an SR method tries to predict the top-K items, with which users will be most likely to interact later.
In this paper, the user set is denoted by $U$ and item set can be represented as
$V$, where $|U|=n$ and $|V|=m$. For each user $u$, the behavior sequence is denoted as $S_u=[v_1,v_2,\ldots,v_{c_u}]$, where $c_u$ is the length of the sequence and $v_i$ is the item interact with user at time $i$. We show the main notations used in this work in Table \ref{notations}.

\begin{table}[]
\caption{The key mathematical notations used in this article}
\label{notations}
\begin{tabular}{cc}
\toprule
Notation & Description \\
\midrule
 $\mathbb{}{B}^n$  & a Poincar\'{e} space of dimension $n$ \\
 $\mathcal{M}$               & Riemannian manifold  \\  
 $c$               & the curvature of Poincar\'{e} space \\
 $\mathcal{T}_\mathbf{x}\mathbb{B}^n$               & the tangent space at point $\mathbf{x}$ with dimension $n$ \\
 $x_v$             &   a item embedding in the Euclid space  \\
\midrule
 $U$  & all users in the training set\\
 $V$ & the item set consisting of all candidate items \\
 $S_u$ & a behavior sequence containing chronological items\\
 $G_r=(V_r,E_r)$  &  the local graph converted from user behavior sequence \\
 $G_G=(V_G,E_G,R_G)$ & the global graph converted from all behavior sequence $S$ \\
 \midrule
 $\mathbf{x}_i^\mathbb{B}$               &    a item embedding in the Poincar\'{e} space  \\
 $L$               &    the layer of graph neural network    \\
 $\omega$ & the magnitude of the auxiliary loss \\

\bottomrule
\end{tabular}
\end{table}

\subsection{Local and Global Graph Construction}
By modeling sequential patterns over adjacent items in each user's behavior sequence, the local graph aims to capture the local-level behavior coherent representation. For each user, behavior sequence $S=[v_1,v_2,...,v_{c_u}]$ is converted into a corresponding local graph $G_L=(V_L,E_L)$, where $V_L \in V$ and $E_L$ denotes the weighted directed edges set composed of $(e^l_{i,i+1})$. Indeed, $e^l_{i,i+1}$ represents the weighted directed edge between clicked item $v_i$ and $v_{i+1}$ at timestamps $i$ and $i+1$, respectively. Since several items may appear in the sequence repeatedly, we assign each edge with a normalized weight, which is calculated as the occurrence of the edge divided by the out-degree of that edge’s start node.

\begin{figure}[ht] 
\centering 
\includegraphics[width=0.8\linewidth]{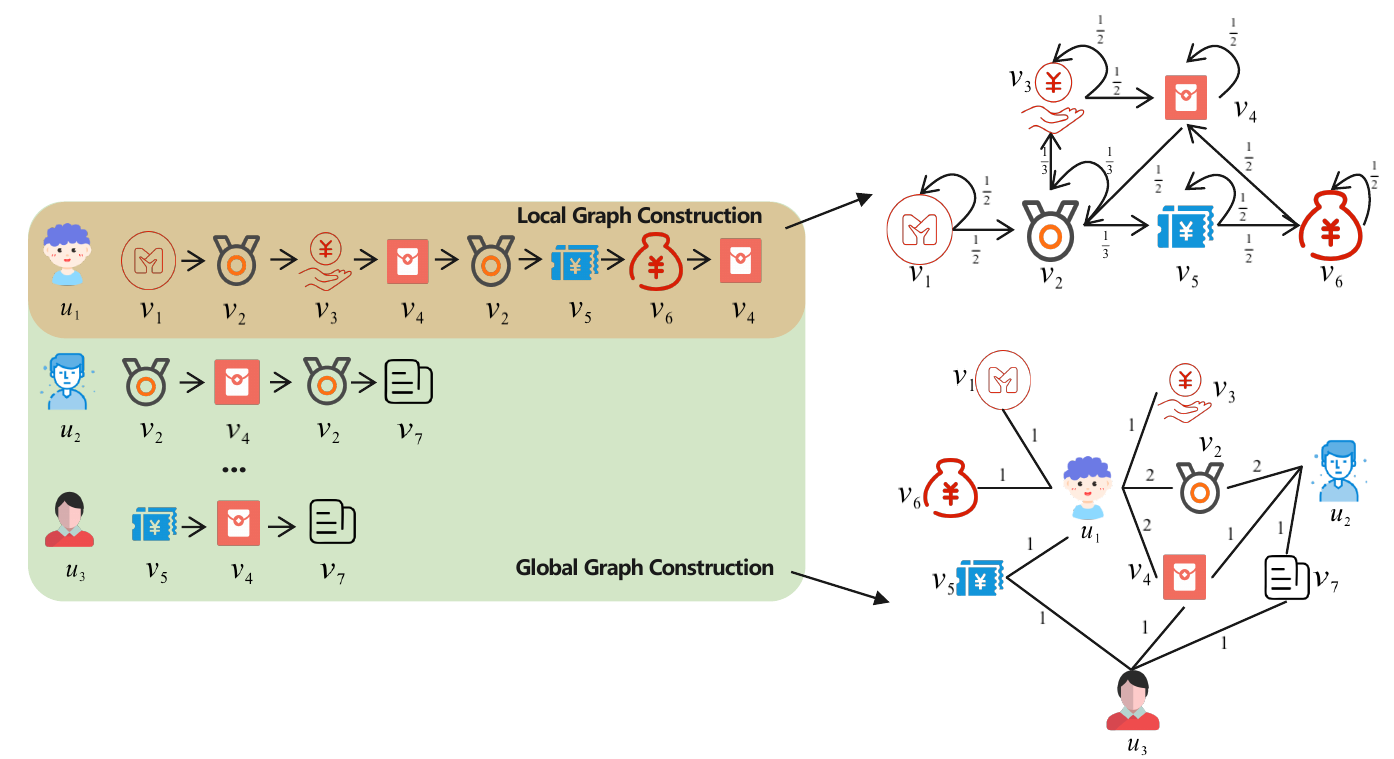}
\caption{The illustration of constructing local and global graph} 
\label{global-graph-consturction} 
\end{figure}

Apart from the local graph construction, we also intend to capture correlated behavior information from a global view for each user. To achieve this, we construct a content associated user-item heterogeneous weighted graph \cite{zhang2019heterogeneous} denoted as $G_G = (V_G,E_G)$ by allocating all the user-item interactions, where $V_G$ denotes the node set that contains user and item entities $V {\cup} U $. $E_G$ denotes the corresponding weighted edges set composed of $(e^g_{i,j})$, where $e^g_{i,j}$ represents the weighted edge between user and corresponding interactive item. Additionally, weight $e^g_{i,j}$ is defined as the number of items $i_j$, which interacted with the user $u_i$. The detailed processes of local and global graph construction are illustrated in Figure \ref{global-graph-consturction}.

\subsection{Hyperbolic Geometry of Poincar\'{e} Ball}
The hyperbolic space $\mathbb{H}^n$ is an $n$-dimensional Riemannian manifold with constant negative curvature. 
Formally, the Poincar\'{e} ball $(\mathbb{B}_c^n,g_{\mathbf{u}}^{\mathbb{B}_c^n})$ with radius $1/\sqrt{c}$ is defined by
\[
\mathbb{B}_c^n=\{\mathbf{u}\in\mathbb{R}^n: c\|\mathbf{u}\|^2<1, c>0\}.
\]
The Riemannian metric tensor is given by
\[
g_{\mathbf{u}}^{\mathbb{B}_c^n}=(\lambda_{\mathbf{u}}^c)^2g_{\mathbf{u}}^\mathbb{E},
\]
where $\lambda_{\mathbf{u}}^c=\frac{2}{1-c\|\mathbf{u}\|^2}$ is the conformal
factor and $g_{\mathbf{u}}^\mathbb{E}=\mathbf{I}_n$ denotes the Euclidean metric tensor. 

Let $\mathcal{T}_{\mathbf{u}}$ denote the tangent space operator. For any  $\mathbf{u},\mathbf{v}\in\mathbb{B}_c^n$ and $\mathbf{x}\in\mathcal{T}_{\mathbf{u}}\mathbb{B}_c^n$, the exponential map $\exp_{\mathbf{u}}^c:\mathcal{T}_{\mathbf{u}}\mathbb{B}_c^n\rightarrow \mathbb{B}_c^n$ and the logarithmic map $\log_{\mathbf{u}}^c:\mathbb{B}_c^n\rightarrow\mathcal{T}_{\mathbf{u}}\mathbb{B}_c^n$ are defined as
\begin{equation}\label{exp}
    \exp_{\mathbf{u}}^c(\mathbf{x})=\mathbf{u}\oplus_c\Bigg(\tanh\Big(\sqrt{c}\frac{\|\mathbf{x}\|\lambda_{\mathbf{u}}^c}{2}\Big)\frac{\mathbf{x}}{\sqrt{c}\|\mathbf{x}\|}\Bigg),
\end{equation}
\begin{equation}\label{log}
    \log_{\mathbf{u}}^c(\mathbf{v})=\frac{2}{\sqrt{c}\lambda_{\mathbf{u}}^c}\tanh^{-1}(\sqrt{c}\|-\mathbf{u}\oplus_c\mathbf{v}\|)\frac{-\mathbf{u}\oplus_c\mathbf{v}}{\|-\mathbf{u}\oplus_c\mathbf{v}\|},
\end{equation}
where $\oplus_c$ is the M\"{o}bius addition defined by
\begin{equation}
\mathbf{u}\oplus_c\mathbf{v}=\frac{(1+2c\langle\mathbf{u},\mathbf{v}\rangle+c\|\mathbf{v}\|^2)\mathbf{u}+(1-c\|\mathbf{u}\|^2)\mathbf{v}}{1+2c\langle\mathbf{u},\mathbf{v}\rangle+c^2\|\mathbf{u}\|^2\|\mathbf{v}\|^2}.
\end{equation}
If $\mathbf{u}=\mathbf{0}$, Eqs. (\ref{exp}) and (\ref{log}) have more appealing forms, i.e., for $\mathbf{x}\in\mathcal{T}_{\mathbf{0}}\mathbb{B}_c^n\setminus \{\mathbf{0}\}$, 
$\mathbf{v}\in\mathbb{B}_c^n\setminus \{\mathbf{0}\}$,
\begin{equation}\label{exp-1}
    \exp_{\mathbf{0}}^c(\mathbf{x})=\tanh(\sqrt{c}\|\mathbf{x}\|)\frac{\mathbf{x}}{\sqrt{c}\|\mathbf{x}\|},
\end{equation}
\begin{equation}\label{log-1}
    \log_{\mathbf{0}}^c(\mathbf{v})=\tanh^{-1}(\sqrt{c}\|\mathbf{v}\|)\frac{\mathbf{v}}{\sqrt{c}\|\mathbf{v}\|}.
\end{equation}
For any $\mathbf{u},\mathbf{v}\in\mathbb{B}_c^n$, the distance is defined by
\begin{equation}\label{dis}
    d_{\mathbb{B}_c^n}(\mathbf{u},\mathbf{v})=\frac{2}{\sqrt{c}}\tanh^{-1}(\sqrt{c}\|-\mathbf{u}\oplus_c\mathbf{v}\|).
\end{equation}

For applying the Poincar\'{e} ball to neural network layers, one additional definition needs to be introduced. For any matrix $\mathbf{A}\in\mathbb{R}^{m\times n}$ and vector $\mathbf{x}\in\mathbb{B}_c^n$, the M\"{o}bius matrix-vector multiplication $\otimes_c~(\mathbb{B}_c^n \rightarrow \mathbb{B}_c^m)$ is introduced as
\begin{equation}
\mathbf{A}\otimes_c\mathbf{u}=\exp_{\mathbf{0}}^c(\mathbf{A}\log_{\mathbf{0}}^c(\mathbf{u})),
\end{equation}
where ${\mathbf{0}}$ is the origin of $\mathbb{R}^n$.

\section{Methodology}
\begin{figure*}[ht]
\centering 
\includegraphics[width=1.0\textwidth,height=1.8in]{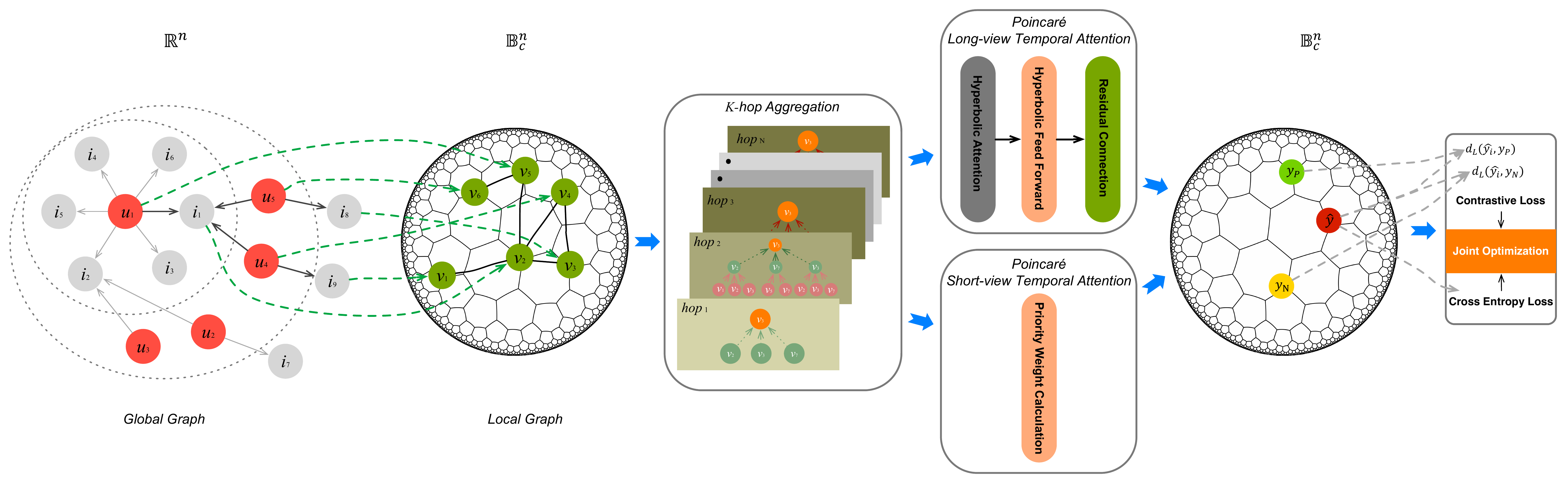}
\caption{The architecture of the PHGR framework. From left to right, all users' historical sequential behaviors are allocated and transformed into a global heterogeneous graph. Besides, for each user, the corresponding local homogeneous graph is constructed by aligning his/her behavior sequence. Then the two graphs are both projected into Poincar\'{e} ball hyperbolic space and following graph aggregation as well as attention operation are conducted. Finally, the predicted probability of the next item that the user most likely to click is obtained.}
\label{PHGR_framework} 
\end{figure*}

In this section, we present our proposed method PHGR as illustrated in Figure \ref{PHGR_framework}.

\subsection{Proposed Inner Product Operation in Poincar\'{e} Ball} \label{Inner Product}

In this subsection, we mainly introduce a new hyperbolic matching score calculation approach, i.e., inner product calculation approach, to alleviate the approximation error caused by the commonly used two-stage method.

Firstly, we recall the definition of inner product in Euclidean space. Let $\mathbf{x},\mathbf{y}\in\mathbb{R}^n$, the Euclidean inner product $\langle\mathbf{x},\mathbf{y}\rangle$ can be defined as follows:
\begin{equation}\label{el-1}
    \langle\mathbf{x},\mathbf{y}\rangle=\sum_{i=1}^n x_iy_i.
\end{equation}
Equivalently, Eq. (\ref{el-1}) can be written as 
\begin{equation}\label{el-2}
    \langle\mathbf{x},\mathbf{y}\rangle=\|\mathbf{x}\|\|\mathbf{y}\|\cos\theta=d_{\mathbb{R}^n}(\mathbf{0},\mathbf{x})d_{\mathbb{R}^n}(\mathbf{0},\mathbf{y})\cos\theta,
\end{equation}
where $\theta$ is the angle between $\mathbf{x}$ and $\mathbf{y}$, or
\begin{equation}
    \begin{split}\label{el-3}
       \langle\mathbf{x},\mathbf{y}\rangle &= \frac{1}{2}(
    \|\mathbf{x}\|^2+\|\mathbf{y}\|^2-\|\mathbf{x}-\mathbf{y}\|^2
    )\\
        & = \frac{1}{2}(
        d_{\mathbb{R}^n}^2(\mathbf{0},\mathbf{x})+d_{\mathbb{R}^n}^2(\mathbf{0},\mathbf{y})-d_{\mathbb{R}^n}^2(\mathbf{x},\mathbf{y})
        ).
    \end{split}
\end{equation}

For calculating inner product in Poincar\'{e} ball hyperbolic space, current works generally apply the two-stage approach, i.e., projection stage and the inner product calculation stage. Specifically, for any $\mathbf{u},\mathbf{v}\in\mathbb{B}^n_c$, the projected inner product $P(\mathbf{u},\mathbf{v})$ is computed by 
\begin{equation}
   P(\mathbf{u},\mathbf{v})=\langle\log_{\mathbf{0}}^c(\mathbf{u}),\log_{\mathbf{0}}^c(\mathbf{v})\rangle
\end{equation}
or
\begin{equation}
    P(\mathbf{u},\mathbf{v})=d_{\mathbb{B}^n_c}(\mathbf{0},\mathbf{u})d_{\mathbb{B}^n_c}(\mathbf{0},\mathbf{v})\cos\beta,
\label{project_inner}
\end{equation}
where $\beta$ is the angle between $\mathbf{u}$ and $\mathbf{v}$ \cite{chamberlain2017neural}.

Referring to the definitions of inner product in Euclidean space (Eqs. (\ref{el-2}) and (\ref{el-3})), we derive the following Theorem 4.1 by omitting the coefficient in Eq. (\ref{dis}) and rewriting it as
\begin{equation}
     d_{\mathbb{B}_c^n}(\mathbf{x},\mathbf{y})=\frac{1}{\sqrt{c}}\tanh^{-1}(\sqrt{c}\|
-\mathbf{x}\oplus_c\mathbf{y}\|).
\label{distance_ou}
\end{equation}

\begin{theorem}\label{thm1}
For any $\mathbf{u},\mathbf{v}\in\mathbb{B}^n_c$, we have 
\begin{equation}\label{rel-1}
    P(\mathbf{u},\mathbf{v})=\langle\log_{\mathbf{0}}^c(\mathbf{u}),\log_{\mathbf{0}}^c(\mathbf{v})\rangle
    =d_{\mathbb{B}^n_c}(\mathbf{0},\mathbf{u})d_{\mathbb{B}^n_c}(\mathbf{0},\mathbf{v})\cos\beta,
\end{equation}
\begin{equation}\label{rel-2}
    P(\mathbf{u},\mathbf{v})\leq
    \frac{1}{2}(
        d_{\mathbb{B}^n_c}^2(\mathbf{0},\mathbf{u})+d_{\mathbb{B}^n_c}^2(\mathbf{0},\mathbf{v})-d_{\mathbb{B}^n_c}^2(\mathbf{u},\mathbf{v})
        ).
\end{equation}
\end{theorem}
\begin{proof}
Following the definition of Euclidean inner product, we can infer that
\begin{equation}
    \langle\log_{\mathbf{0}}^c(\mathbf{u}),\log_{\mathbf{0}}^c(\mathbf{v})\rangle
=\|\log_{\mathbf{0}}^c(\mathbf{u})\|\|\log_{\mathbf{0}}^c(\mathbf{v})\|\cos\beta^{'},
\end{equation}
where $\beta^{'}$ is the angle between $\log_{\mathbf{0}}^c(\mathbf{u})$ and $\log_{\mathbf{0}}^c(\mathbf{v})$. Based on Eq. (\ref{distance_ou}), the following equation could be easily derived:
\begin{equation}
    \|\log_{\mathbf{0}}^c(\mathbf{u})\|=d_{\mathbb{B}^n_c}(\mathbf{0},\mathbf{u}), 
\label{logu_du}
\end{equation}
\begin{equation}
    \|\log_{\mathbf{0}}^c(\mathbf{v})\|=d_{\mathbb{B}^n_c}(\mathbf{0},\mathbf{v}).
\label{logv_dv}
\end{equation}
Then if we can further infer $\cos\beta^{'}$ $=$ $\cos\beta$, the Eq. (\ref{rel-1}) could be proved. Since the Poincar\'{e} ball is conformal to Euclidean space, the angle between two vectors $\mathbf{u}$, $\mathbf{v}$ is given by
\begin{equation}
    \cos\beta^{'}=\frac{\langle\log_{\mathbf{0}}^c(\mathbf{u}),\log_{\mathbf{0}}^c(\mathbf{v})\rangle}{\|\log_{\mathbf{0}}^c(\mathbf{u})\|\|\log_{\mathbf{0}}^c(\mathbf{u})\|}=\frac{\langle\tanh^{-1}(
\sqrt{c}\|\mathbf{u}\|)\frac{\mathbf{u}}{\sqrt{c}\|\mathbf{u}\|},\tanh^{-1}(
\sqrt{c}\|\mathbf{v}\|)\frac{\mathbf{v}}{\sqrt{c}\|\mathbf{v}\|}\rangle}{\|\tanh^{-1}(
\sqrt{c}\|\mathbf{u}\|)\frac{\mathbf{u}}{\sqrt{c}\|\mathbf{u}\|}\|\|\tanh^{-1}(
\sqrt{c}\|\mathbf{v}\|)\frac{\mathbf{v}}{\sqrt{c}\|\mathbf{v}\|}\|}=\frac{\langle\mathbf{u},\mathbf{v}\rangle}{\|\mathbf{u}\|\|\mathbf{v}\|}=\cos\beta.
\end{equation}
Therefore, we can prove Eq. (\ref{rel-1}). 

For proving Eq. (\ref{rel-2}), since
\begin{equation}
    P(\mathbf{u},\mathbf{v})=\frac{1}{2}(\|\log_{\mathbf{0}}^c(\mathbf{u})\|^2+\|\mathbf{\log_{\mathbf{0}}^c(\mathbf{v})}\|^2-\|\mathbf{\log_{\mathbf{0}}^c(\mathbf{u})}-\mathbf{\log_{\mathbf{0}}^c(\mathbf{v})}\|^2),
\end{equation}
based on Eqs. (\ref{logu_du}) and (\ref{logv_dv}), we only need to prove that 
\begin{equation}
    d_{\mathbb{B}^n_c}(\mathbf{u},\mathbf{v})\geq \|\log_{\mathbf{0}}^c(\mathbf{u})-\log_{\mathbf{0}}^c(\mathbf{v})\|.
\label{distance_gap}
\end{equation}
Let hyperbolic curvature $c=1$, and $\beta$ be the angle between vectors $\mathbf{u}$ and $\mathbf{v}$. Referring to Figure \ref{gap1}, the top of figure denotes the hyperbolic space and the bottom shows its projection to the origin in Euclidean space. Analogously, $OA=\mathbf{u}$, $OB=\mathbf{v}$, then
\begin{equation}
    \|\log_{\mathbf{0}}^c(\mathbf{u})-\log_{\mathbf{0}}^c(\mathbf{v})\|=\|A'B'\|.
\end{equation}
Therefore, the inequality (\ref{distance_gap}) can be rewritten as 
\begin{equation}
    d_{\mathbb{B}^n_c}(A,B) \geq \|A'B'\|.
\label{distance_gap_2}
\end{equation}
Let $r$ be the hyperbolic length of $OA$:
\begin{equation}
    r=d_{\mathbb{B}^n_c}(O,A)=\|\log_{\mathbf{0}}^c(\mathbf{u})\|=\|O'A'\|.
\end{equation}
For $O'A'$, $O'B'$, $A'B'$, they satisfy the trigonometric formulae in Euclidean space, i.e., 
\begin{equation}
    \frac{\|O'A'\|}{\sin \angle O'B'A'}=\frac{\|O'B'\|}{\sin \angle O'A'B'}=\frac{\|A'B'\|}{\sin \angle A'O'B'},
\end{equation}
where $\|O'A'\|$, $\|O'B'\|$, $\|A'B'\|$ are the lengths of the sides of a triangle, and $\angle O'B'A'$, $\angle O'A'B'$, and $\angle A'O'B'$ denote the corresponding opposite angles. Then we can get 
\begin{equation}
  \|A'B'\|=\sin \beta \frac{\|O'A'\|}{\sin \angle O'B'A'}=\sin \beta \frac{r}{\sin \angle O'B'A'}.
\label{euclid_tri}
\end{equation}
Analogously, $OA$, $OB$, $\overset{\frown}{AB}$ satisfy the hyperbolic trigonometric formulae, i.e.,
\begin{equation}
    \frac{\sinh (d_{\mathbb{B}^n_c}(O,A))}{\sin \angle OBA}=\frac{\sinh (d_{\mathbb{B}^n_c}(O,B))}{\sin \angle OAB}=\frac{\sinh (d_{\mathbb{B}^n_c}(A,B))}{\sin \angle AOB},
\end{equation}
where $d_{\mathbb{B}^n_c}(O,A)$, $d_{\mathbb{B}^n_c}(O,B)$, and $d_{\mathbb{B}^n_c}(A,B)$ represent the hyperbolic lengths of the sides of a triangle, and $\angle OBA$, $\angle OAB$, and $\angle AOB$ are the corresponding opposite angles under hyperbolic space. Then we can infer that
\begin{equation}
    \sinh (d_{\mathbb{B}^n_c}(A,B))=\sin\beta \frac{\sinh (d_{\mathbb{B}^n_c}(O,A))}{\sin \angle OBA}=\sin\beta \frac{\sinh r}{\sin \angle OBA}.
\label{hypolic_tri}
\end{equation}
Based on the fact that $\sinh k$ is monotonically increasing when $k\in [0,+\infty)$ and $\sinh 0=0$, we can get
\begin{equation}
    \sinh(\eta k)=\sinh(\eta k+(1-\eta)*0)\leq \eta\sinh k+(1-\eta)\sinh 0=\eta\sinh k
\end{equation}
if $0\leq \eta \leq 1$. Since $0\leq \angle OBA \leq \angle O'B'A' \leq \frac{\pi}{2}$, as
\begin{equation}
    \frac{\sinh r}{\sin \angle OBA}\geq\frac{\sinh r}{\sin \angle O'B'A'}.
\end{equation}
So based on Eqs. (\ref{euclid_tri}) and (\ref{hypolic_tri}) we can infer that:
\begin{equation}
     \sinh (d_{\mathbb{B}^n_c}(A,B))=\sin(\beta) \frac{\sinh r}{\sin \angle OBA}\geq \sin(\beta) \frac{\sinh r}{\sin \angle O'B'A'} \geq \sinh(\sin \beta \frac{r}{\sin \angle O'B'A'})=\sinh(\|A'B'\|).
\end{equation}
Therefore, we can get $d_{\mathbb{B}^n_c}(A,B) \geq \|A'B'\|$, which concludes the proof of the inequality (\ref{distance_gap_2}).

\end{proof}

\begin{figure}[ht] 
\centering 
\includegraphics[width=0.5\linewidth]{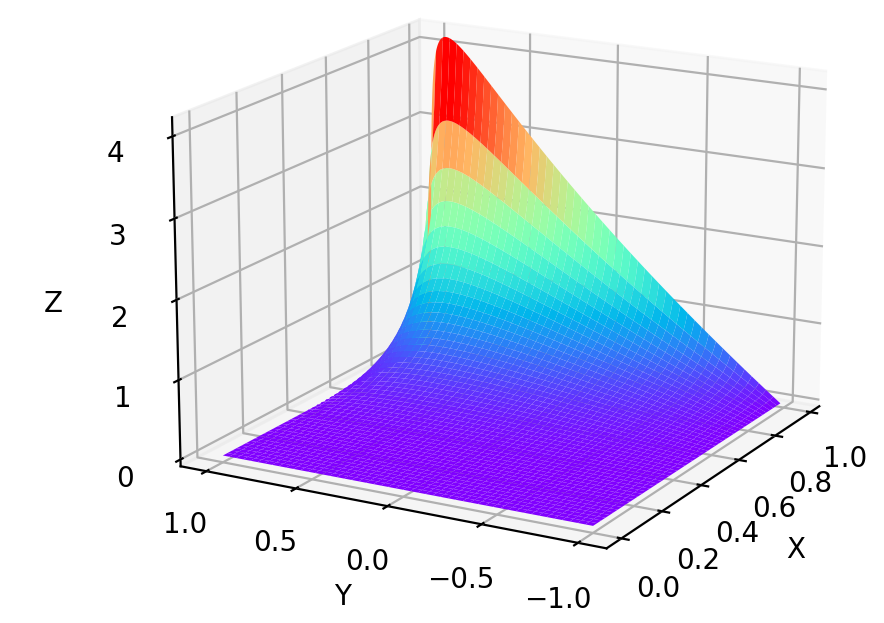}
\caption{The difference of $P(\mathbf{u},\mathbf{v})$ and $D(\mathbf{u},\mathbf{v})$, where we let $\|\mathbf{u}\|=\|\mathbf{v}\|$. The X axis represents $\|\mathbf{u}\|$, the Y axis represents $\cos\beta$, where $\beta$ is the angle between $\mathbf{u}$ and $\mathbf{v}$, and the Z axis represents the difference.} 
\label{distance} 
\end{figure}

Based on the above discussion, we propose a novel computation mechanism of the inner product in Poincar\'{e} ball space based on the geodesic distance. Indeed, the present new hyperbolic operator $D(\mathbf{u},\mathbf{v})$ computing the inner product in Poincar\'{e} ball is defined as follows:
\begin{equation}
 D(\mathbf{u},\mathbf{v})=\frac{1}{2}(d_{\mathbb{B}_c^n}^2(\mathbf{0},\mathbf{u})
  +d_{\mathbb{B}_c^n}^2(\mathbf{0},\mathbf{v})-d_{\mathbb{B}_c^n}^2(\mathbf{u},\mathbf{v})).
\label{new_product}
\end{equation}


In the following part, we discuss the advantage of the newly proposed hyperbolic inner product in Eq. (\ref{new_product}) comparing with the commonly used projected hyperbolic inner product in Eq. (\ref{project_inner}). Let us use an example to give an elaborate illustration. As shown in Figure \ref{gap1}, let $\angle AOB=\angle A_1OB_1<\frac{\pi}{2}$ and $\| OA\|\| OB\| < \|  OA_1\|  \|  OB_1\| $. Besides, we set $\| OA\|=\| OB\|$ and $\| OA_1\|=\| OB_1\|$ for calculation simplicity. By the definition of $P(\textbf{u},\textbf{v})=d_{\mathbb{B}_c^n}(\textbf{0},\textbf{u})d_{\mathbb{B}_c^n}(\textbf{0},\textbf{v})\cos\beta$, we can 
always conclude that $P(OA,OB)<P(OA_1,OB_1)$, which is in accord with the intuition in Euclidean space as $\| OA\| < \|  OA_1\|$ and thus $<A_1, B_1>$ is more similar, i.e., gets higher matching score, than that of $<A, B>$. But as shown in Figure \ref{gap1}, $<A_1, B_1>$ is more dissimilar than that of $<A, B>$ as $A_1$ and $B_1$ get farther from the origin and the distance between $A_1$ and $B_1$ increases exponentially with radius under hyperbolic space. By the definition of our proposed calculation method, we can prove that $D(OA_1,OB_1)<P(OA_1,OB_1)$ as Eq. (\ref{rel-2}). Figure \ref{distance} shows the numerical difference of $D(\mathbf{u},\mathbf{v})$ and $P(\mathbf{u},\mathbf{v})$, in which we can see that $P(\mathbf{u},\mathbf{v})$ is always equal or greater than $D(\mathbf{u},\mathbf{v})$. The equality holds if and only if $\cos \beta =\pm 1$.

\begin{figure}[ht]
\centering
\subfigure[]{
\begin{minipage}[t]{0.5\linewidth}
\centering
\includegraphics[width=8.2cm]{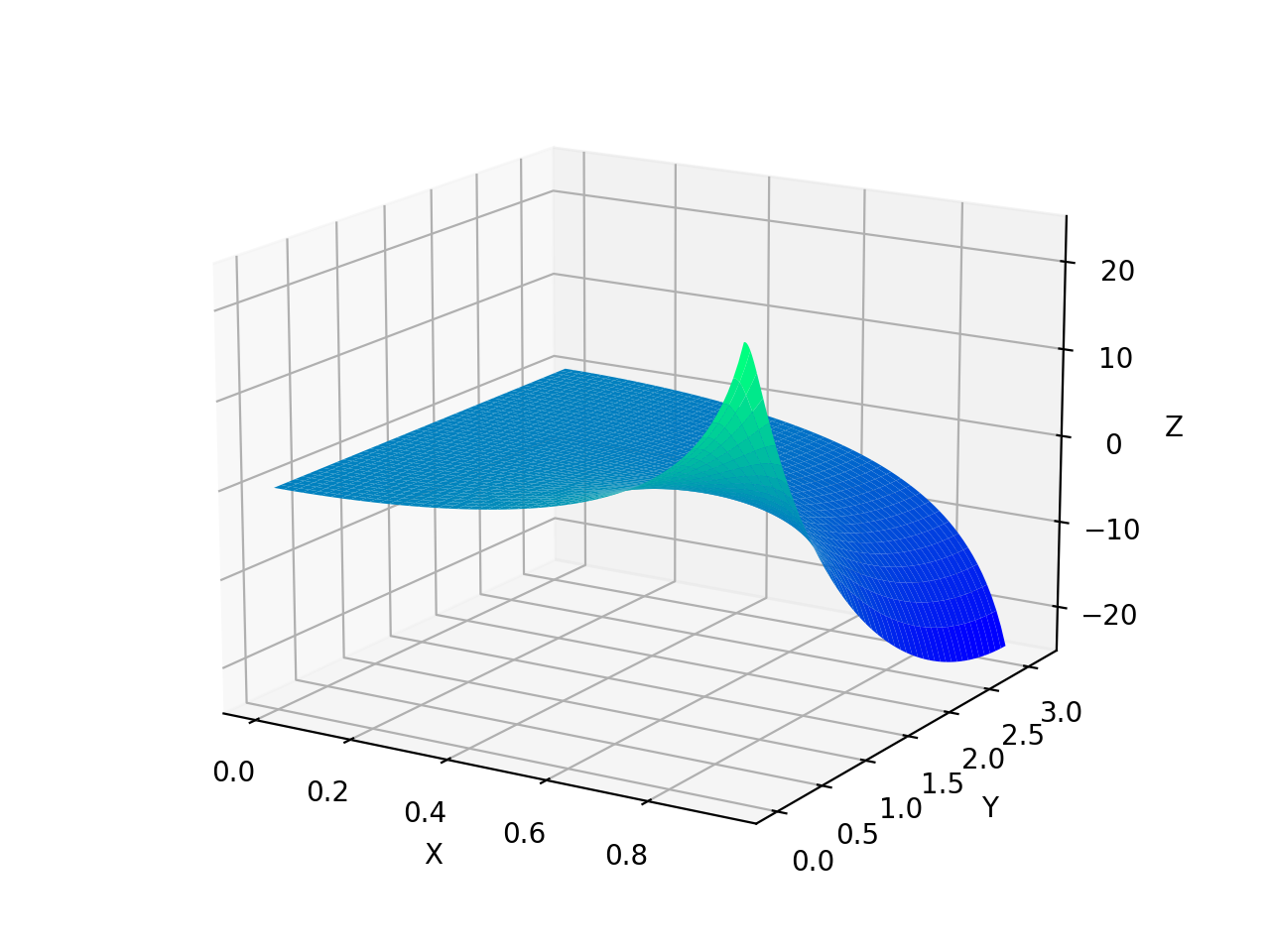}
\end{minipage}%
}%
\subfigure[]{
\begin{minipage}[t]{0.5\linewidth}
\centering
\includegraphics[width=7cm]{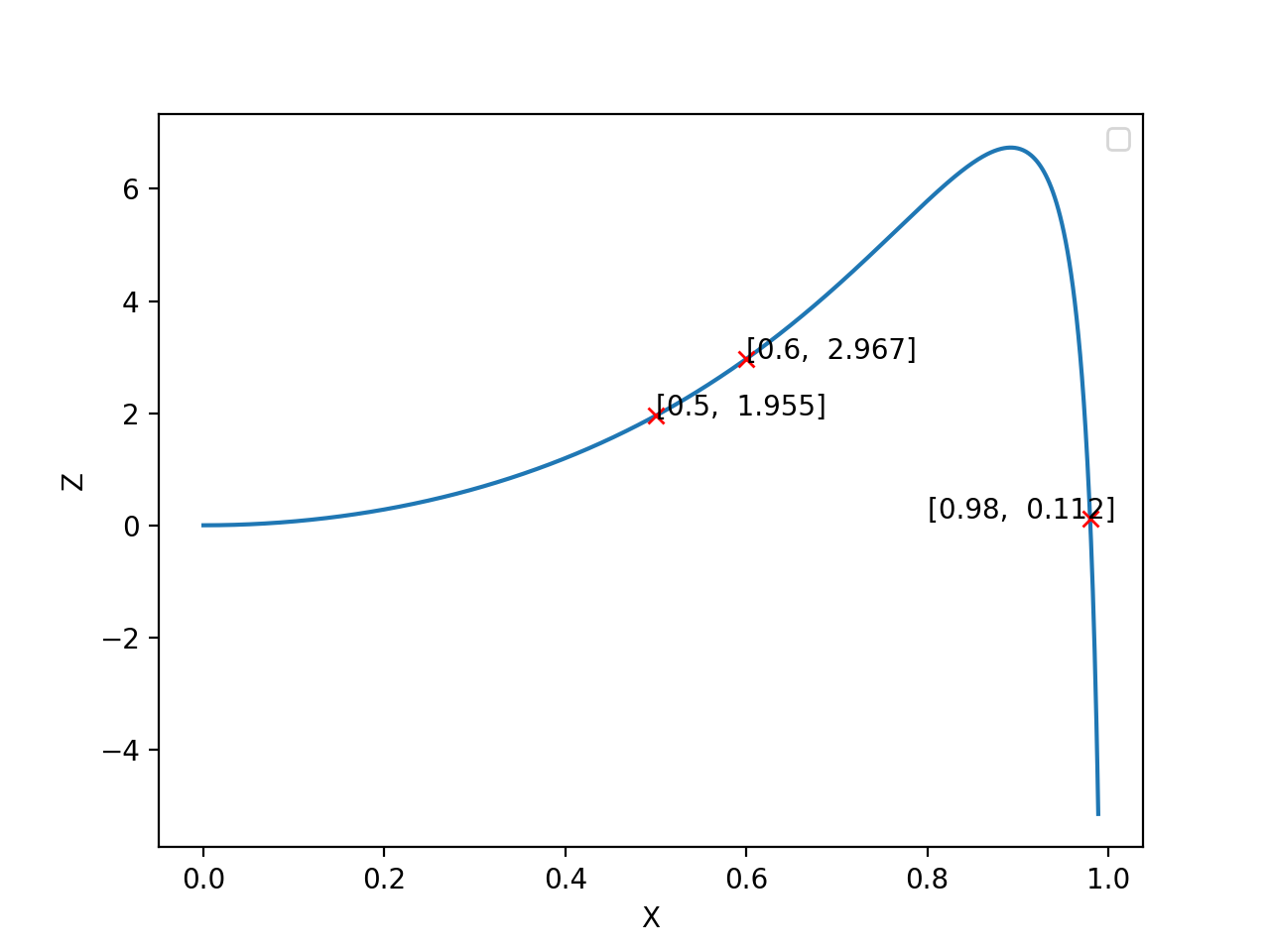}
\end{minipage}%
}%
\centering
\caption{The illustration of the mathematical characteristic of the proposed hyperbolic inner product calculation method. (a).The variation tendency of the calculated inner product with the variation of angle $\beta$ and $\|\mathbf{u}\|$ (we set $\|\mathbf{u}\|=\|\mathbf{v}\|$ for simplicity). Specifically, the X axis represents $\|\mathbf{u}\|$, the Y axis represents the angle $\beta$ between $\mathbf{u}$ and $\mathbf{v}$, and the Z axis represents the calculated inner product in hyperbolic space. (b).The variation tendency of calculated inner product with the variation of $\|\mathbf{u}\|$ under the condition of fixed angle $\beta$, which is a slice from Figure 5(a).}
\label{product_angle_norm}
\end{figure}

However, the magnitude relationship of $D(OA,OB)$ and $D(OA_1,OB_1)$ is not permanently changeless. The mathematical characteristic of the proposed hyperbolic inner product operation is shown in Figure \ref{product_angle_norm}. Specifically, based on Figure \ref{product_angle_norm}(a), we can infer that the variation tendency of the calculated hyperbolic inner product with the variation of angle $\beta$ and $\|\mathbf{u}\|$ (we set $\|\mathbf{u}\|=\|\mathbf{v}\|$ for simplicity) is not monotonous. Indeed, the inner product $D(\textbf{u},\textbf{v})$ increase at the early stage and then decrease as the increasing of the $\|\mathbf{u}\|$ under the condition of fixed angle $\beta$ as illustrated in Figure \ref{product_angle_norm}(b). For example, if we set $\| OA\|$ = 0.5 and $\| OA_1\|$ = 0.6, then $D(OA,OB)<D(OA_1,OB_1)$. But if we set $\| OA\|$ = 0.5 and $\| OA_1\|$ = 0.98, then $D(OA,OB)>D(OA_1,OB_1)$. Thus, such characteristics could alleviate the issue caused by the commonly used projected hyperbolic inner product as Eq (\ref{project_inner}). Moreover, the new inner product operation  can make the training more stable because it keeps $\|\mathbf{u}\|$ from the boundary of Poincar\'{e} ball.

\subsection{Graph Learning Under Poincar\'{e} ball Space}
Before conducting global and local graph representation learning under hyperbolic space, we first project the user and item node within corresponding global and local graph into Poincar\'{e} ball. Specifically, users and items are initially embedded by sampling the Gaussian distribution in Euclidean space and then we project them to Poincar\'{e} ball using exponential map $\exp_{\mathbf{0}}^c$. The projected user and item embeddings can be represented by    
\begin{equation}
\mathbf{x}_u^{\mathbb{B}}= \exp_{\mathbf{0}}^c(\mathbf{x}_u),
\end{equation}
\begin{equation}
\mathbf{x}_i^{\mathbb{B}}=\exp_{\mathbf{0}}^c(\mathbf{x}_i),x_u,x_i \in \mathbb{R}^d,
\end{equation}
where $x_u$, $x_i$ denotes user and item initial embeddings in Euclidean space, respectively.

\subsubsection{Poincar\'{e} Global-level Graph Attention Network}
\label{global-graph-learn}
As illustrated in Figure \ref{PHGR_framework}, We impose a shared user-item heterogeneous graph by aligning all users and items from provided dataset to directly encode low-order interactions from a global view then complement the local homogeneous graph convolution. The reason is that local item-related homogeneous graph cannot aggregate the direct correlations between user-item pairs. 

Referring to the traditional graph aggregation operation under Euclidean space, we adopt graph attention mechanism during the iterative information aggregating process between target node and its corresponding neighbors under Poincar\'{e} ball space. Indeed, the attention weights for one target item node and its surrounding user neighbors can be formulated as follow: 
\begin{equation}
\label{eq15}
e^G_{iu} =\delta_{u\in \mathcal{N^G}(i)}(\mathbf{W_{iu}}(\log_\mathbf{0}^c(\mathbf{x}_{i}^{\mathbb{B},G,l})||\log_\mathbf{0}^c(\mathbf{x}_u^{\mathbb{B},G,l}))+{b_{iu}}),
\end{equation}
where $\mathbf{W}_{iu}\in\mathbb{R}^{2d}$ and $b_{iu}$ are learnable parameter. $\delta$ represents the softmax function.  ${x}_u^{\mathbb{B},G,l}$ and ${x}_i^{\mathbb{B},G,l}$ denote the user embedding and item embedding obtained at $l$-th global graph representation layer in Poincar\'{e} ball space, and $\mathbf{x}_{i}^{\mathbb{B},G,0}=\mathbf{x}_{i}^{\mathbb{B}}$. $\mathcal{N^G}(i)$ represents the neighbour set of node $i$ within the global graph.
Then the hyperbolic global-level item representations could be iteratively updated with these attention coefficients as follow:
\begin{equation}
\label{eq16}
\mathbf{x}_i^{\mathbb{B},G,l+1} = \exp_{\mathbf{x}_i^{\mathbb{B},G,l}}^c(\sum_{u \in \mathcal{N^G}(i)}e_{iu}\log_{\mathbf{x}_i^{\mathbb{B},G,l}}^c(\mathbf{x}_u^{\mathbb{B},G,l})),
\end{equation}
After stacking multiple Poincar\'{e} graph attention layers to fully aggregate information from higher-order relationships, the hyperbolic item representation within the global user-item graph could be obtained as:
\begin{equation}
\label{eq17}
\mathbf{\mathbf{x}}_i^{\mathbb{B},G} = \exp_\mathbf{0}^{c}(\sum_{l=0}(\alpha_l\ast\log_\mathbf{0}^{c}(\mathbf{x}_i^{\mathbb{B},G,l}) )),
\end{equation} 
where $\alpha_{l}$ is the empirical parameters to control the magnitude of high-order connection information. The hyperbolic global-level user representations are updated in the similar way and not presented for simplicity. 
\subsubsection{Poincar\'{e} Local-level Graph Attention Network}
\label{local-graph-learn}
As illustrated in Figure \ref{global-graph-consturction}, for each user, his/her sequential historical behaviors are converted into a directed weighted graph to extract preference from a local view as such information still explicitly under-explored by the global user-item heterogeneous graph. Referring to the framework shown in Figure \ref{PHGR_framework}, the outputs of the global user-item graph representation learning are used to enhance the local item-item graph modeling. The local graph attention convolution operations are similar to section \ref{global-graph-learn}. Specifically, the attention weights between two neighbor items can be formulated as:
\begin{equation}
\label{eq19}
e^L_{ij} =\delta_{j \in \mathcal{N^L}(i)}(\mathbf{W_{ij}}(\log_\mathbf{0}^c(\mathbf{x}_{i}^{\mathbb{B},L,l})||\log_\mathbf{0}^c(\mathbf{x}_{i}^{\mathbb{B},L,l}))+{b_{ij}}),
\end{equation}
where $\mathbf{W_{ij}}\in\mathbb{R}^{2d}$ and $b_{ij}$ are learnable parameter. ${x}_i^{\mathbb{B},L,l}$ denotes the item embedding obtained at $l$-th local graph layer in Poincar\'{e} ball space. $\mathcal{N^L}(i)$ represents the neighbour node of $i$ within local graph, and $\mathbf{x}_{i}^{\mathbb{B},L,0}=\mathbf{x}_{i}^{\mathbb{B},G}$.
Then the hyperbolic local-level item representations could be iteratively updated through
\begin{equation}
\label{eq20}
\mathbf{x}_{i}^{\mathbb{B},L,l+1} = \exp_{\mathbf{x}_i^{\mathbb{B},L,l}}^c(\sum_{j \in \mathcal{N^L}(i)}e^L_{ij}\log_{\mathbf{x}_{i}^{\mathbb{B},L,l}}^c(\mathbf{x}_{j}^{\mathbb{B},L,l})),
\end{equation}
The out of Poincar\'{e} local graph attention network is
\begin{equation}
\label{eq21}
\mathbf{x}_i^{\mathbb{B},L} = \exp_\mathbf{0}^{c}(\sum_{l=0}(\zeta_l\ast\log_\mathbf{0}^{c}(\mathbf{x}_{i}^{\mathbb{B},L,l}) )),
\end{equation}
where $\zeta_l$ is empirical parameters to control the magnitude of different $l$-th learned representations.
\subsection{Temporal Attention Mechanism}
After global and local graph convolution, the item representations $\mathbf{X}^{\mathbb{B},L}= [\mathbf{x}_{1}^{\mathbb{B},L},\mathbf{x}_{2}^{\mathbb{B},L},\ldots,\mathbf{x}_{n}^{\mathbb{B},L}]$ are obtained. For each user, to capture sequential patterns within his/her historical behaviors, we further conduct attention operation among the temporal behaviors as such temporal information is explicitly under-explored by the global and local graph representation learning. Specifically, we conduct two attention operations, i.e., long-view temporal attention and short-view temporal attention, based on each user's sequential interacted items as illustrated in Figure \ref{PHGR_framework}.

Following the commonly used attention mechanism \cite{vaswani2017attention} under Euclidean space, the hyperbolic self-attention operation among one user's sequential interacted items are re-formalized as follows:
\begin{equation}
\label{eq22}
\begin{split}
\mathbf{X}^{\mathbb{B},long} =&\delta(\frac{(\mathbf{W}_Q\log_\mathbf{0}^c(\mathbf{X}^{\mathbb{B},L}))
(\mathbf{W}_K\log_\mathbf{0}^c(\mathbf{X}^{\mathbb{B},L}))^\top}{\sqrt{d}})(\mathbf{W}_V\log_\mathbf{0}^c(\mathbf{X}^{\mathbb{B},L})),
\end{split}
\end{equation}
where $\mathbf{W}_Q$, $\mathbf{W}_K$, $\mathbf{W}_V \in \mathbb{R}^{d \times d}$ are learnable parameter. Then we can get the sequential patterns information from a long-view through:    
\begin{equation}
\label{eq23}
\mathbf{z}^{\mathbb{B},long}=\exp_\mathbf{0}^{c}(\frac{\sum_{i=1}^n \mathbf{x}_i^{\mathbb{B},LT}}{n}).
\end{equation}

Besides, for one user, the next most likely behavior is often related to his/her nearest interests, thus sequential patterns information from a short-term view is also crucial for predicting user preference. Specifically, the nearest item $n$ acts as query and historical interacted items $[1,2,\ldots,n-1]$ act as keys as well as values. The attention weight between item $n$ and item $i$ is formulated as follows:
\begin{equation}
\label{eq24}
\gamma_{i}=\mathbf{q}^{T}\sigma(\mathbf{W}_{n}\log_\mathbf{0}^c({\mathbf{x}_{n}^{\mathbb{B},L}})+\mathbf{W}_{i}\log_\mathbf{0}^c({\mathbf{x}_{i}^{\mathbb{B},L}})),
\end{equation}
where $\mathbf{W}_{n}, \mathbf{W}_{i} \in \mathbb{R}^{d\times d}$ and $\mathbf{q}^{T} \in \mathbb{R}^d$ are learnable parameters. Then we can get the sequential patterns information from a short-view through:    
\begin{equation}
\label{eq25}
\mathbf{z}^{\mathbb{B},short}=\exp_\mathbf{0}^{c}(\sum_{i=1}^n(\gamma_{i}\ast\log_\mathbf{0}^c(\mathbf{x}_{i}^{\mathbb{B},L}))),
\end{equation}.

\subsection{Model Training and Prediction}
As illustrated in Figure \ref{PHGR_framework}, through the graph representation learning and temporal attention operation under hyperbolic space, the final user representation could be obtained as follows:

\begin{equation}
\label{eq26}
\mathbf{{x}_{u}^{\mathbb{B}}}=\mathbf{W}[ \log_\mathbf{0}^c(\mathbf{x}_{n}^{\mathbb{B},L}) || \log_\mathbf{0}^c(\mathbf{z}^{\mathbb{B},long}) || \log_\mathbf{0}^c(\mathbf{z}^{\mathbb{B},short})],
\end{equation}
where $\mathbf{W} \in \mathbb{R}^{d \times (3\ast d)}$ is a learnable parameter. The unified preference probability between user $u \in U$ and item $v_i \in V$ is given as:
\begin{equation}
\widehat{\mathbf{y}_i}=\delta(D(\mathbf{{x}_{u}^{\mathbb{B}}},\mathbf{x}_{i}^{\mathbb{B}})),
\end{equation}
where $D(\mathbf{{x}_{u}^{\mathbb{B}}},\mathbf{x}_{i}^{\mathbb{B}})$ follows Eq. (\ref{new_product}).
Inspired by the work \cite{guo2021hcgr}, we incorporate main and auxiliary tasks to train our model. The loss of main task is cross-entropy $L_{ce} $, which has been widely applied in SR task, can be defined by
\begin{equation}					
L_{ce}=-\sum_{i=1}^{m}{({\mathbf{y}_i}\log{(\widehat{\mathbf{y}_i})}+(1-{\mathbf{y}_i})\log{(1-\widehat{\mathbf{y}_i})})},
\end{equation}
the loss of auxiliary task is the contrastive ranking loss $L_{cr}$ given by
\begin{equation}
L_{cr}=\sum_{i=1}^{m}{\max((d_L(\widehat{\mathbf{y}_i},\mathbf{y}_P)-d_L(\widehat{\mathbf{y}_i},\mathbf{y}_N)+\xi,0)},
\end{equation}
and the overall loss function can be defined as
\begin{equation}
\label{loss}
L_{total}= L_{ce}+\omega\ast L_{cr}.
\end{equation}
where $\omega$ controls the magnitude of the auxiliary loss $L_{cr}$. 


\section{Experiment}
In this section, we first present the experimental settings including datasets, comparison methods, evaluation metrics and parameter settings.
Then, we conduct comprehensive experiments on three widely used real-world datasets and one newly collected industrial dataset to answer the following questions:
\begin{itemize}
\item{
Q1: How does PHGR perform in SR scenarios comparing with the state-of-the-art methods?
}
\item{
Q2: How each component of PHGR contributes to the performance?}
\item{
Q3: How is the effectiveness and portability of the Poincar\'{e} ball module of PHGR?}
\end{itemize}
\subsection{Experimental Settings} \label{Experimental Settings}
\subsubsection{Datasets Description}
We provide a summary of the four representative real-world datasets of our experiments in Table \ref{tab:data}. Specifically, the first three datasets are widely-used {Amazon datasets}\footnote{\url{http://jmcauley.ucsd.edu/data/amazon/}}, each of which refers to a top-level category of products, including beauty (Beauty), pet supplies (Pet), and tools \& home improvements (T\&H), respectively. The MYbank dataset is a larger-scaled and more challenging dataset collected from Ant Group, which describes users' interactions in financial products such as debit, trust, and accounting. For each dataset, we randomly choose 80$\%$, 10$\%$, and 10$\%$ of the sequences as training, validation, and testing datasets. And we apply early stopping with patience of 10, which means we stop training if the loss does not decrease for 10 consecutive epochs. 

\begin{table}[H]

\centering
\caption{Dataset descriptions.}
\label{tab:data}
\resizebox{0.5\linewidth}{!}{ 
\begin{tabular}{l|lllr} 
\toprule
Dataset      & Beauty  & Pet    & TH     & MYbank    \\ 
\hline
\#Users      & 109,725 & 78,875 & 94,596  & 691,701   \\
\#Items      & 45,908  & 23,894 & 40,688 & 3,188     \\
\#Avg.I/User & 5.22   & 5.00  & 4.96  & 8.38     \\
\#Avg.U/Item & 12.48  & 16.50 & 11.53 & 1818.01  \\
\#Actions    & 0.57M   & 0.47M  & 0.39M  & 5.79~M    \\
\bottomrule
\end{tabular}
}
\end{table}

\subsubsection{Comparison Methods}
We compare PHGR with the following SR methods:
\begin{itemize}
\item \textbf{FPMC} \cite{rendle2010factorizing} - a classical Markov-chain-based method, which considers the latest interaction.
\item \textbf{FOSSIL} \cite{he2016fusing} - a classical Markov-chain-based method, which captures personalized dynamics.
\item \textbf{GRU4Rec} \cite{hidasi2015session} - a representative RNN-based method, which stacks multiple GRU layers for session-parallel mini-batch training.
\item \textbf{NARM} \cite{li2017neural} - a hybrid encoder with attention mechanism to model sequential behaviors.
\item \textbf{HGN} \cite{ma2019hierarchical} -  a hierarchical gating network integrated with the Bayesian personalized ranking for SR.
\item \textbf{SASRec} \cite{kang2018self} - an attention-based sequential method, which utilizes few actions and considers long-range dependencies.
\item \textbf{LightSANs} \cite{fan2021lighter} - a low-rank decomposed self-attention network.
\item \textbf{HME} \cite{feng2020hme} - a hyperbolic metric embedding method, which exploits two bipartite graphs to learn the representations of users and POIs.
\item \textbf{SRGNN} \cite{wu2019session} – a graph-based method to learn item representations.
\item \textbf{GC-SAN} \cite{xu2019graph} – an improved version of SRGNN to compute sequence-level embeddings.
\item \textbf{LESSR} \cite{chen2020handling} – a session-based method with GNN, which utilizes auxiliary graph to generate item representation.
\end{itemize}

\subsubsection{Evaluation Metrics}
We employ three widely used metrics, i.e., $HitRate (H), NDCG (N), MAP (M)$, to evaluate the performance of SR methods. Here, we choose $K$ to be 10 and 20 to show the different metrics for $H@K$, $N@K$ and $M@K$. For all the three metrics, higher values indicate better performance.
\subsubsection{Parameter Settings}
The hyperparameters of our proposed PHGR are tuned on the validation set by grid search. Specifically, the learning rate is tuned among $[10^{-4}, 10^{-3}]$, the embedding size $d$ is tuned from the range of $\{8, 16, 32, 64, 128\}$, and the number of aggregation layer $L$ is tuned among $[1, 5]$. We set batch-size $b$ to be 128, curvature $k$ to be 1, and the magnitude of the auxiliary loss $\omega$ is tuned among $[10^{-4}, 10^0]$. The hyperparameters of all comparison methods are set according to their original papers or found by grid research. In the following section, we will investigate the impact of key hyperparameters in greater depth.


\subsection{Performance on SR (for Q1)}
\begin{figure}[H]
\centering

\subfigure[Results on Beauty.]{
\begin{minipage}[t]{0.35\linewidth}
\centering
\includegraphics[width=2.3in]{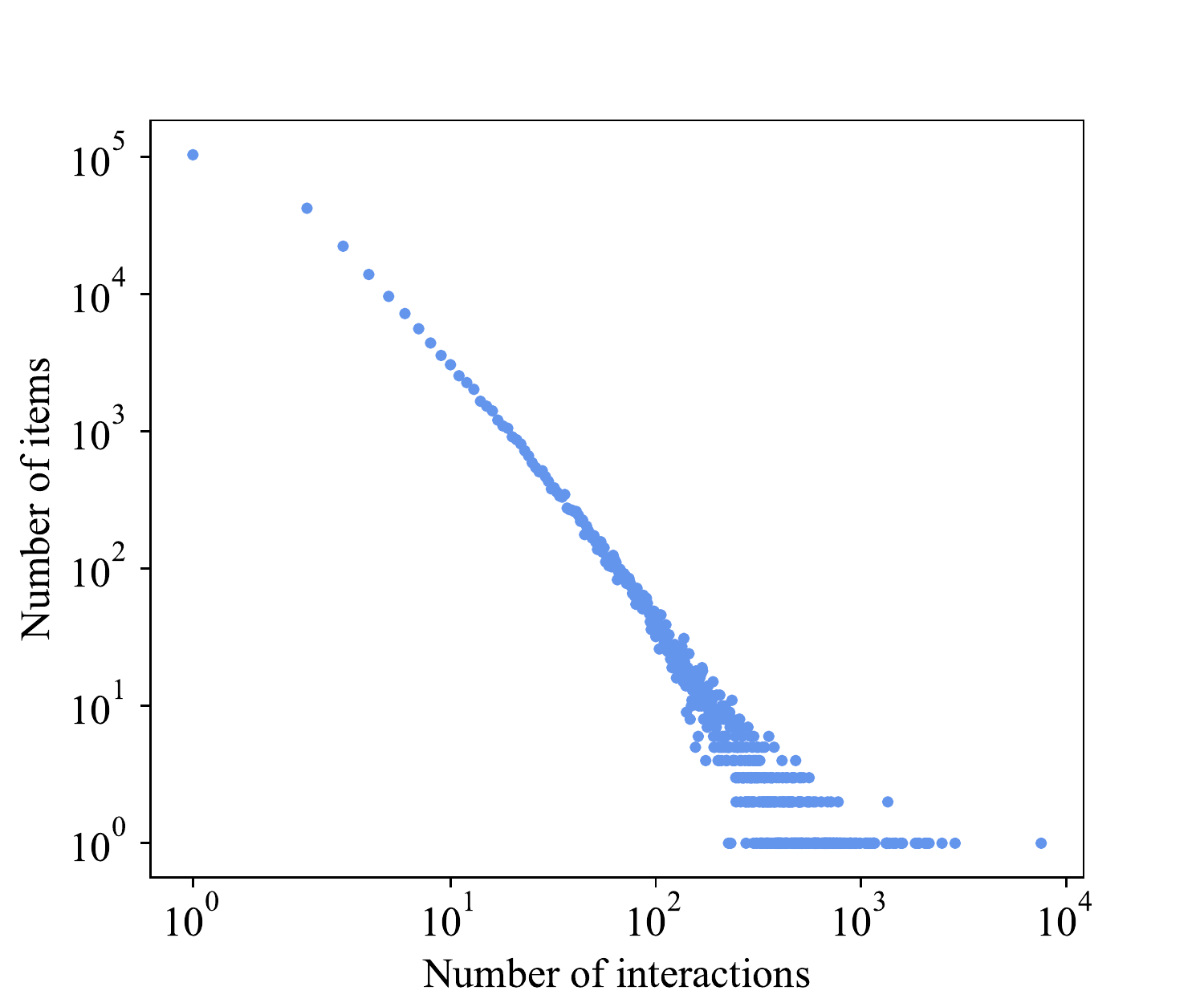}
\end{minipage}%
}%
\subfigure[Results on Pet.]{
\begin{minipage}[t]{0.35\linewidth}
\centering
\includegraphics[width=2.3in]{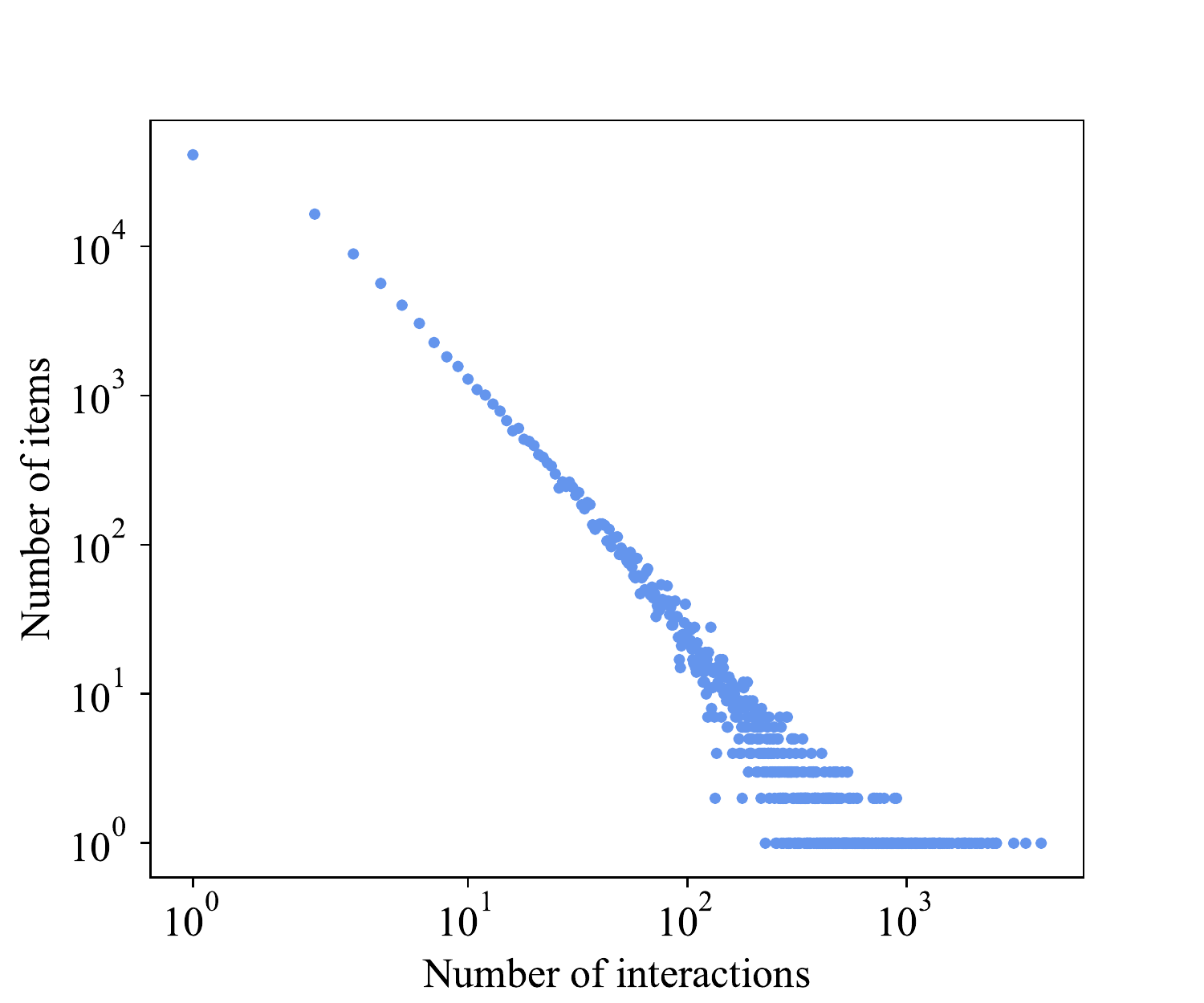}
\end{minipage}%
}%

\subfigure[Results on TH.]{
\begin{minipage}[t]{0.35\linewidth}
\centering
\includegraphics[width=2.3in]{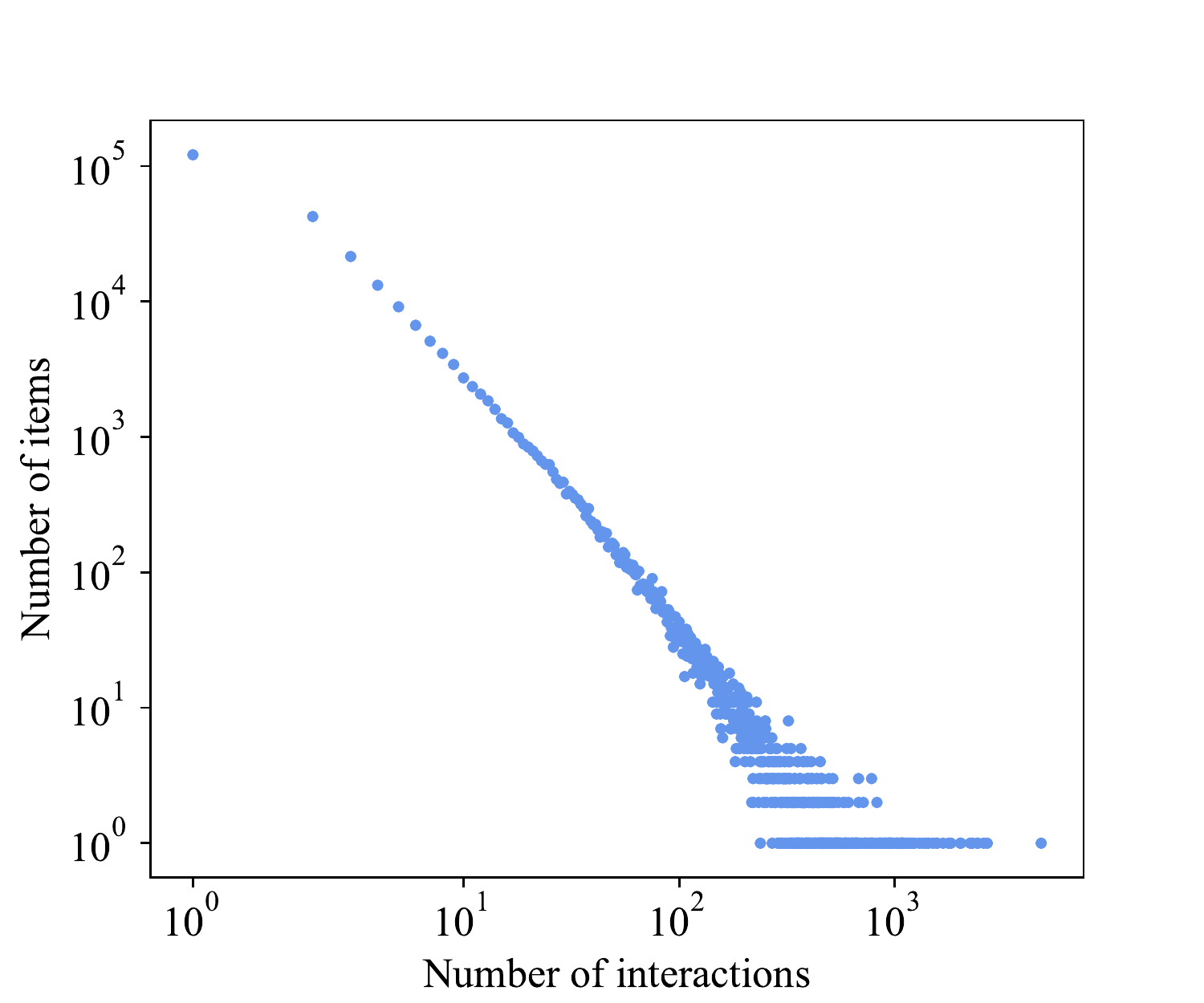}
\end{minipage}
}%
\subfigure[Results on MYbank.]{
\begin{minipage}[t]{0.35\linewidth}
\centering
\includegraphics[width=2.3in]{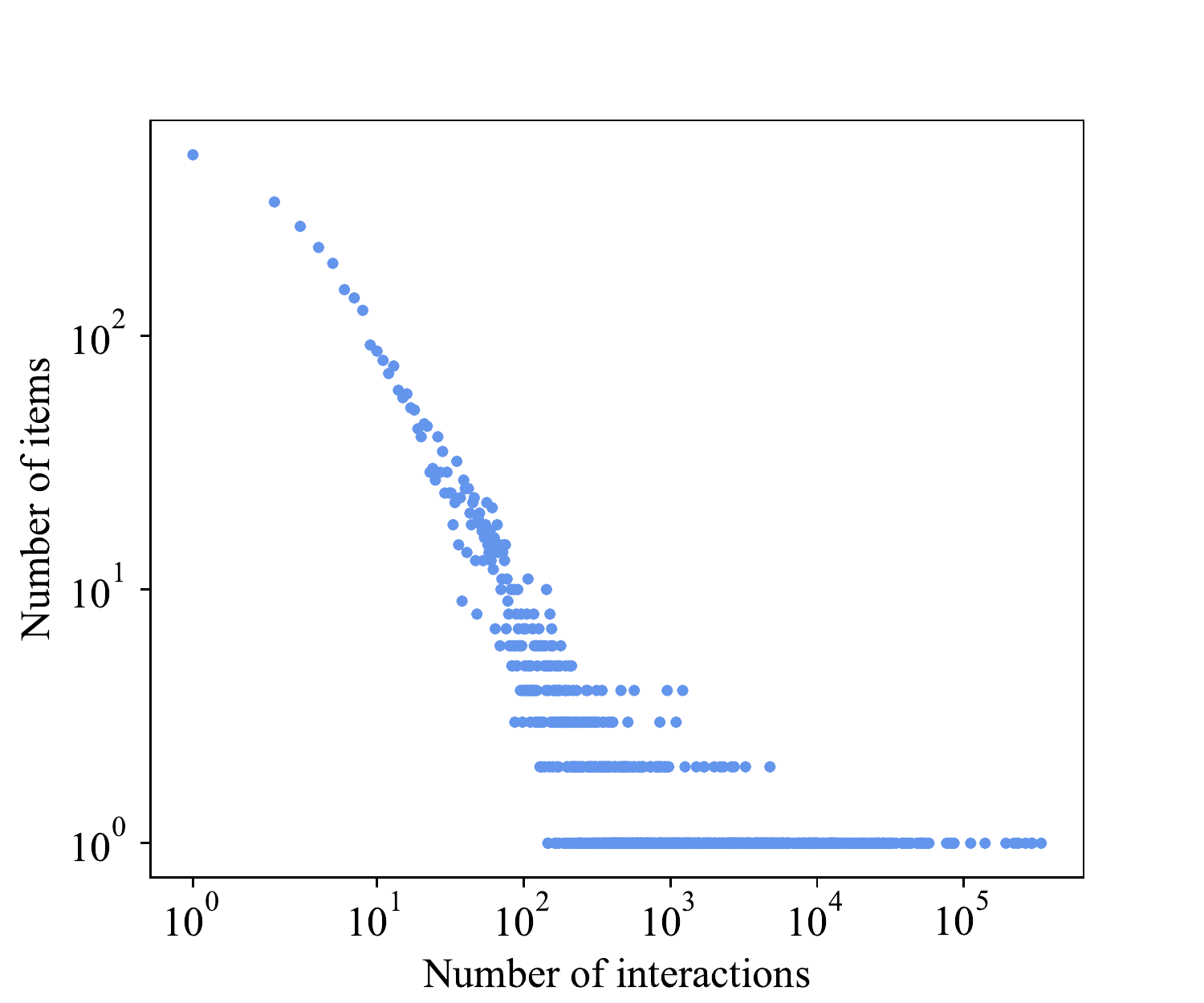}
\end{minipage}
}%
\centering
\caption{Clicking distributions of items in each of the datasets. The X-axis presents the number of interactions associated with a user or item, and the Y-axis shows the number of such users or items.
\label{fig:clicking_distribution}
}
\end{figure}

To answer Q1, we present the HitRate, NDCG, and MAP values in Table \ref{tab:exp1}. The highest value under each metric is highlighted in boldface and the second-highest value is underlined. 
Furthermore, to gain a deeper insight into PHGR's characteristics, we present the distributions of user-item interactions on each of the datasets in Figure \ref{fig:clicking_distribution} and investigate how PHGR performs under various data distributions. 

\begin{table}[H]
\centering
\caption{Performance illustration of all comparison methods on four datasets.}
\label{tab:exp1}
\resizebox{0.97\linewidth}{!}{ 

\begin{tabular}{c|c|ccccccccccc|c|c}
\hline
\multirow{2}{*}{Dataset} & \multirow{2}{*}{Metric} & \multicolumn{11}{c|}{Comparision Methods}                                                                              & Proposed Methods & \multirow{2}{*}{Improve(\%)} \\ \cline{3-14}
                         &                         & FPMC  & FOSSIL & GRU4Rec & NARM       & HGN   & SASRec     & LightSANs & HME   & SRGNN      & GCSAN       & LESSR      & PHGR             &                              \\ \hline
\multirow{6}{*}{Beauty}  & H@10                    & 4.69  & 5.03   & 4.89    & 5.03       & 4.88  & 1.47       & 1.19      & 5.80  & 5.09       & 6.00        & \underline{6.13} & \textbf{6.62}    & 7.93                         \\
                         & N@10                    & 2.76  & 2.33   & 2.92    & 3.02       & 2.83  & 0.70       & 0.59      & 3.23  & \underline{3.35} & 3.08        & 2.94       & \textbf{3.69}    & 10.15                        \\
                         & M@10                    & 2.16  & 1.50   & 2.33    & 2.41       & 2.19  & 0.47       & 0.41      & 2.40  & \underline{2.81} & 2.18        & 1.95       & \textbf{2.87}    & 1.96                         \\
                         & H@20                    & 6.22  & 6.93   & 6.59    & 6.80       & 6.62  & 2.23       & 1.78      & 7.77  & 6.39       & 7.98        & \underline{8.34} & 8.54             & 2.38                         \\
                         & N@20                    & 3.14  & 2.81   & 3.35    & 3.47       & 3.26  & 0.89       & 0.74      & 3.65  & \underline{3.67} & 3.58        & 3.50       & \textbf{4.19}    & 14.24                        \\
                         & M@20                    & 2.26  & 1.63   & 2.44    & 2.53       & 2.31  & 0.52       & 0.45      & 2.51  & \underline{2.90} & 2.31        & 2.10       & \textbf{2.97}    & 2.31                         \\ \hline
\multirow{6}{*}{Pet}     & H@10                    & 4.65  & 5.46   & 5.22    & 5.58       & 5.72  & 6.49       & 5.55      & 4.63  & 4.72       & \underline{6.58}  & 6.34       & \textbf{6.84}    & 3.91                         \\
                         & N@10                    & 2.44  & 2.56   & 2.87    & 3.20       & 3.16  & \underline{3.28} & 3.01      & 4.25  & 2.79       & 3.25        & 2.98       & \textbf{3.77}    & 14.85                        \\
                         & M@10                    & 1.76  & 1.66   & 2.16    & \underline{2.47} & 2.38  & 2.29       & 2.24      & 2.61  & 2.20       & 2.22        & 1.95       & \textbf{2.85}    & 15.32                        \\
                         & H@20                    & 6.73  & 7.78   & 7.73    & 7.87       & 8.31  & 9.14       & 8.40      & 6.30  & 6.84       & 9.16        & \underline{9.16} & \textbf{9.49}    & 3.64                         \\
                         & N@20                    & 2.96  & 3.14   & 3.50    & 3.77       & 3.81  & \underline{3.95} & 3.72      & 5.30  & 3.32       & 3.90        & 3.69       & \textbf{4.45}    & 12.63                        \\
                         & M@20                    & 1.90  & 1.82   & 2.33    & \underline{2.63} & 2.56  & 2.48       & 2.43      & 2.90  & 2.34       & 2.40        & 2.14       & \textbf{3.04}    & 15.41                        \\ \hline
\multirow{6}{*}{TH}      & H@10                    & 3.72  & 3.39   & 2.78    & 3.25       & 3.53  & 2.02       & 2.56      & 4.09  & 2.83       & 4.46        & \underline{4.60} & \textbf{4.75}    & 3.29                         \\
                         & N@10                    & 2.24  & 1.64   & 1.6     & 1.90       & 2.01  & 1.06       & 1.42      & 2.23  & 1.78       & 2.30        & \underline{2.56} & \textbf{2.58}    & 0.75                         \\
                         & M@10                    & 1.79  & 1.10   & 1.24    & 1.49       & 1.55  & 0.78       & 1.07      & 1.67  & 1.46       & 1.62        & \underline{1.93} & \textbf{1.96}    & 1.52                         \\
                         & H@20                    & 4.86  & 4.54   & 3.94    & 4.45       & 4.91  & 3.18       & 3.72      & 5.43  & 3.74       & 5.83        & \underline{6.05} & \textbf{6.35}    & 4.96                         \\
                         & N@20                    & 2.53  & 1.93   & 1.89    & 2.20       & 2.36  & 1.35       & 1.71      & 2.55  & 2.01       & 2.64        & \underline{2.93} & \textbf{2.97}    & 1.36                         \\
                         & M@20                    & 1.87  & 1.18   & 1.32    & 1.57       & 1.65  & 0.86       & 1.15      & 1.84  & 1.52       & 1.72        & \underline{2.03} & \textbf{2.19}    & 7.99                         \\ \hline
\multirow{6}{*}{MYbank}  & H@10                    & 51.35 & 42.40  & 55.6    & 54.64      & 48.41 & 52.13      & 53.06     & 56.03 & 53.22      & \underline{55.88} & 54.72      & \textbf{57.55}   & 2.99                         \\
                         & N@10                    & 34.53 & 28.37  & 37.66   & 37.25      & 30.37 & 35.59      & 35.55     & 38.44 & 36.50      & \underline{37.86} & 37.38      & 38.49            & 1.67                         \\
                         & M@10                    & 29.32 & 24.03  & 32.09   & 31.84      & 24.80 & 30.45      & 30.67     & 31.95 & 31.30      & \underline{32.27} & 32.00      & \textbf{33.10}   & 2.58                         \\
                         & H@20                    & 61.94 & 51.30  & 65.66   & 64.76      & 59.93 & 62.24      & 63.23     & 67.71 & 63.15      & \underline{66.40} & 64.80      & \textbf{67.86}   & 2.20                         \\
                         & N@20                    & 37.22 & 30.62  & 40.20   & 39.81      & 33.29 & 38.15      & 38.12     & 41.10 & 39.01      & \underline{40.53} & 39.93      & \textbf{41.63}   & 2.70                         \\
                         & M@20                    & 30.06 & 24.64  & 32.79   & 32.55      & 25.60 & 31.16      & 31.42     & 33.80 & 31.98      & \underline{33.00} & 32.70      & \textbf{34.12}   & 3.39                         \\ \hline
\end{tabular}
}
\end{table}

\textbf{Analysis}: From Table \ref{tab:exp1}, we can observe that PHGR outperforms all the comparison methods with an average improvement of 5.83\%. 
Specifically, compared with Markov chain-based methods, i.e., FPMC and FOSSIL, PHGR performs better probably since it is capable of capturing the sequential dependencies among items in a sequence, which is crucial for exploiting dynamic user preference in SR scenarios. In addition, PHGR outperforms the RNN-based and transformer-based methods, i.e., GRU4Rec, NARM, SASRec, and LightSANs, because it can model multiple-way transitions between consecutive items and pays attention to the transitions among the contexts, which effectively learning user representations even though user behaviors are sparse.
In comparison to the graph-based methods, i.e., SRGNN, GC-SAN, and LESSR, PHGR shows its superiority in capturing the heterogeneity and hierarchy information contained in the contrusted user-item and item-item graphs, which is equivalent to extract the coarse-to-fine transition patterns from various transition types respectively. 
Furthermore, comparing with hyperbolic-based method, i.e., HME, the performance of PHGR still gets an obvious improvement due to the carefully designed framework for SR scenarios and the contribution of each module will be illustrated in section \ref{ablation}.
Besides, we provide a deeper analysis from the perspective of data structure. In Figure \ref{fig:clicking_distribution}, the distribution of item interactions of four datasets is plotted  and all of them present a long-tailed shape, whose degrees between the most popular items and unpopular ones are quite large. This data structure is also called as hierarchy-like structure.
Considering PHGR's improvements on this type of datasets, it could be implied that PHGR is more adaptive in the sequential recommendation scenarios with hierarchical properties. 
Specifically, we also find an interesting phenomenon that the transformer-based methods (SASRec and LightSANs) perform poorly in the Beauty and TH datasets, even worse than the previous methods. Given that the sparsity of user behaviors in the Beauty and TH datasets as shown in Table \ref{tab:data}, 12.48 and 11.53 for the average number of the users per item, relatively less than the other datasets, transformer-based models fail to learn the multiplex transitions between consecutive items  since they pay much more attention between the interactions.

\subsection{Ablation Studies (for Q2)} 
\label{ablation}
To answer Q2, we conduct ablation experiments with seven simplified versions of PHGR: (1) \textbf{EHGR}: PHGR without the Poincar\'{e} ball module, i.e., the Poincar\'{e} ball module is replaced with the Euclidean space module, (2) \textbf{PHGR-w/o-IP}: PHGR without the Poincar\'{e} inner product module, i.e., the inner product can't be performed on the Poincar\'{e} ball module and can only be performed after the elements are projected onto the Euclidean space, (3) \textbf{PHGR-w/o-G}: PHGR without the global-level graph attention network, and (4) \textbf{PHGR-w/o-L}: PHGR without the local-level graph attention network, (5)  \textbf{PHGR-w/o-Long}: PHGR without long-view temporal attention network,  (6)  \textbf{PHGR-w/o-short}: PHGR without short-view temporal attention work, (7)  \textbf{PHGR-w/o-L\&S}: PHGR without both long and short temporal attention network.
All the ablation results are presented in Table \ref{tab:exp2}.

\begin{table}[H]
\centering
\caption{Performance of ablation studies on four datasets.}
\label{tab:exp2}
\resizebox{0.97\linewidth}{!}{ 
\begin{tabular}{c|c|cccccccc}
\hline
Dataset                 & Metric & PHGR  & EHGR  & PHGR-w/o-IP & PHGR-w/o-G & PHGR-w/o-L & PHGR-w/o-Long & PHGR-w/o-Short & PHGR-w/o-L\&S \\ \hline
\multirow{6}{*}{Beauty} & H@10   & 6.62  & 6.15  & 6.49        & 6.56       & 6.50       & 6.35          & 5.87           & 5.33          \\
                        & N@10   & 3.69  & 2.96  & 3.67        & 3.54       & 3.58       & 3.46          & 3.41           & 3.32          \\
                        & M@10   & 2.87  & 1.98  & 2.82        & 2.61       & 2.68       & 2.66          & 2.58           & 2.46          \\
                        & H@20   & 8.54  & 8.41  & 8.48        & 8.82       & 8.70       & 8.41          & 7.67           & 7.92          \\
                        & N@20   & 4.19  & 3.54  & 4.17        & 4.11       & 4.14       & 3.98          & 3.87           & 3.60          \\
                        & M@20   & 2.97  & 2.11  & 2.93        & 2.76       & 2.83       & 2.78          & 2.72           & 2.56          \\ \hline
\multirow{6}{*}{Pet}    & H@10   & 6.84  & 6.67  & 6.77        & 6.65       & 6.56       & 5.99          & 5.85           & 5.33          \\
                        & N@10   & 3.77  & 3.68  & 3.74        & 3.52       & 3.54       & 3.42          & 3.34           & 3.32          \\
                        & M@10   & 2.85  & 2.76  & 2.81        & 2.56       & 2.61       & 2.69          & 2.53           & 2.46          \\
                        & H@20   & 9.49  & 9.26  & 9.34        & 9.3        & 8.82       & 8.39          & 8.35           & 8.12          \\
                        & N@20   & 4.45  & 4.33  & 4.39        & 4.19       & 4.11       & 4.06          & 3.94           & 3.90          \\
                        & M@20   & 3.04  & 2.94  & 2.99        & 2.75       & 2.76       & 2.86          & 2.69           & 2.66          \\ \hline
\multirow{6}{*}{TH}     & H@10   & 4.75  & 4.59  & 4.68        & 4.65       & 4.09       & 4.46          & 4.25           & 3.85          \\
                        & N@10   & 2.58  & 2.56  & 2.58        & 2.46       & 2.26       & 2.40          & 2.31           & 2.25          \\
                        & M@10   & 1.96  & 1.93  & 1.96        & 1.79       & 1.69       & 1.77          & 1.72           & 1.75          \\
                        & H@20   & 6.35  & 6.04  & 6.23        & 6.23       & 5.51       & 5.99          & 5.65           & 5.06          \\
                        & N@20   & 2.97  & 2.93  & 2.97        & 2.86       & 2.61       & 2.79          & 2.67           & 2.56          \\
                        & M@20   & 2.19  & 2.03  & 2.16        & 1.9        & 1.78       & 1.87          & 1.81           & 1.84          \\ \hline
\multirow{6}{*}{MYbank} & H@10   & 57.55 & 56.86 & 57.00       & 54.05      & 54.56      & 57.25         & 56.95          & 56.55         \\
                        & N@10   & 38.49 & 38.43 & 38.50       & 36.65      & 36.99      & 38.13         & 37.76          & 37.73         \\
                        & M@10   & 33.10 & 32.71 & 32.76       & 31.24      & 31.53      & 32.92         & 32.70          & 32.00         \\
                        & H@20   & 67.86 & 67.44 & 67.64       & 64.67      & 65.27      & 67.16         & 67.14          & 66.59         \\
                        & N@20   & 41.63 & 41.10 & 41.19       & 39.34      & 39.69      & 41.12         & 40.73          & 40.80         \\
                        & M@20   & 34.12 & 33.44 & 33.49       & 31.98      & 32.27      & 33.87         & 33.55          & 33.50         \\ \hline
\end{tabular}
}
\end{table}

\textbf{EHGR}:
EHGR is the version that PHGR replaces the Poincar\'{e} ball module with the Euclidean space module but it still contains the user-item heterogeneous graph representation learning to encode low-order interactions from a global view then complement the local homogeneous item-item graph convolution. The reason of such design is that
local item-related homogeneous graph cannot aggregate the direct correlations between user-item pairs for accurately obtaining user preference. Comparing PHGR with EHGR, we find that PHGR outperforms EHGR with an average improvement of 7.80\% 
on all datasets. Without the Poincar\'{e} ball module, EHGR can't fully guarantee the representation learning operation of hierarchy-like data structure, which leads to degradation of the performance.

\textbf{PHGR-w/o-IP}:
Comparing PHGR with PHGR-w/o-IP, we find that PHGR outperforms PHGR-w/o-IP with an average improvement of 1.01\% 
on all datasets.
The performance drops greatly when removing the proposed inner product operation on most metrics, but slightly obvious in terms of the N@10.
Without the Poincar\'{e} inner product module,  PHGR-w/o-IP needs to project the embedding vectors from the Poincar\'{e} ball space to the Euclidean space when performing an inner product, which leads to the inaccuracies of the distance under the  Poincar\'{e} metric. In other words, the novel designed inner product operation under Poincaré ball could reduce the cumulative
error of general bidirectional projection process between Poincaré ball and Euclidean space and is quite appropriate for calculating hyperbolic inner product considering intrinsic geodesic distance.

\textbf{PHGR-w/o-G}: 
Comparing PHGR with PHGR-w/o-G and PHGR-w/o-L, we find that PHGR outperforms both ablation methods on all datasets as shown in Table \ref{tab:exp2}. Heterogeneous global graph consisting of users and items provides additional information to item-item graph representation learning through an implicit message passing, which complements the user preference from the view of global low-order user-item interactions. This additional information significantly contributes to the model performance as the PHGR-w/o-G gets a poor evaluation. PHGR-w/o-L also behaves badly, since the item information is also inadequate. 

\textbf{PHGR-w/o-L}: The local-level graph is sequence-based and homogeneous, in which, item representation is learned by aggregating neighbor information within a user's clicking history. 
In a conclusion, without global or local-level graph, PHGR is unable to capture the comprehensive user preferences, which could be enhanced from users' direct clicking items or implicitly correlated ones. PHGR improves the model performance by simultaneously modeling both global and local users' interesting.

\textbf{PHGR-w/o-Long}: Long-view temporal attention network is a re-formalized self-attention mechanism  under the hyperbolic space. As observed in Table \ref{tab:exp2}, without this attention network, performance drops compared to the PHGR. It could suggest that the long-view attention operation captures an enhanced context-based user preference by learning the item interactions within the whole clicking sequence, which contributes the next-item prediction.

\textbf{PHGR-w/o-Short}: Short-view temporal attention learns the item interactions focusing on the previous clicked item and the idea is inspired by that user's most likely behavior is generally closer to his/her nearest interests. We can discover in Table \ref{tab:exp2} that the performance of PHGR-w/o-Short is even worse than the PHGR-w/o-Long with most metrics on all datasets. Short-view temporal attention captures users' instant interests through utilizing the item information just before the next clicking, which greatly helps to understand users' future behavior as the experiment result. 

\textbf{PHGR-w/o-L\&S}: From Table \ref{tab:exp2}, we can find that without both long and short attention network, PHGR-w/o-L\&S behaves worst even compared with other ablation model on most metrics and datasets. As discussed above, long and short attention networks provide a context-based information and users' instant interests respectively from item interactions learning in different temporal scale. These attention-based operation further explores the users' preference under a implicitly weighted interactions learning beyond graph-based SR method, which significantly improves model performance as demonstrated.


\subsection{Portability of PHGR's Poincar\'{e} ball Module (for Q3)}
\begin{figure}[H]
\centering
\subfigure[Results on Beauty.]{
\begin{minipage}[t]{0.4\linewidth}
\centering
\includegraphics[width=2.3in]{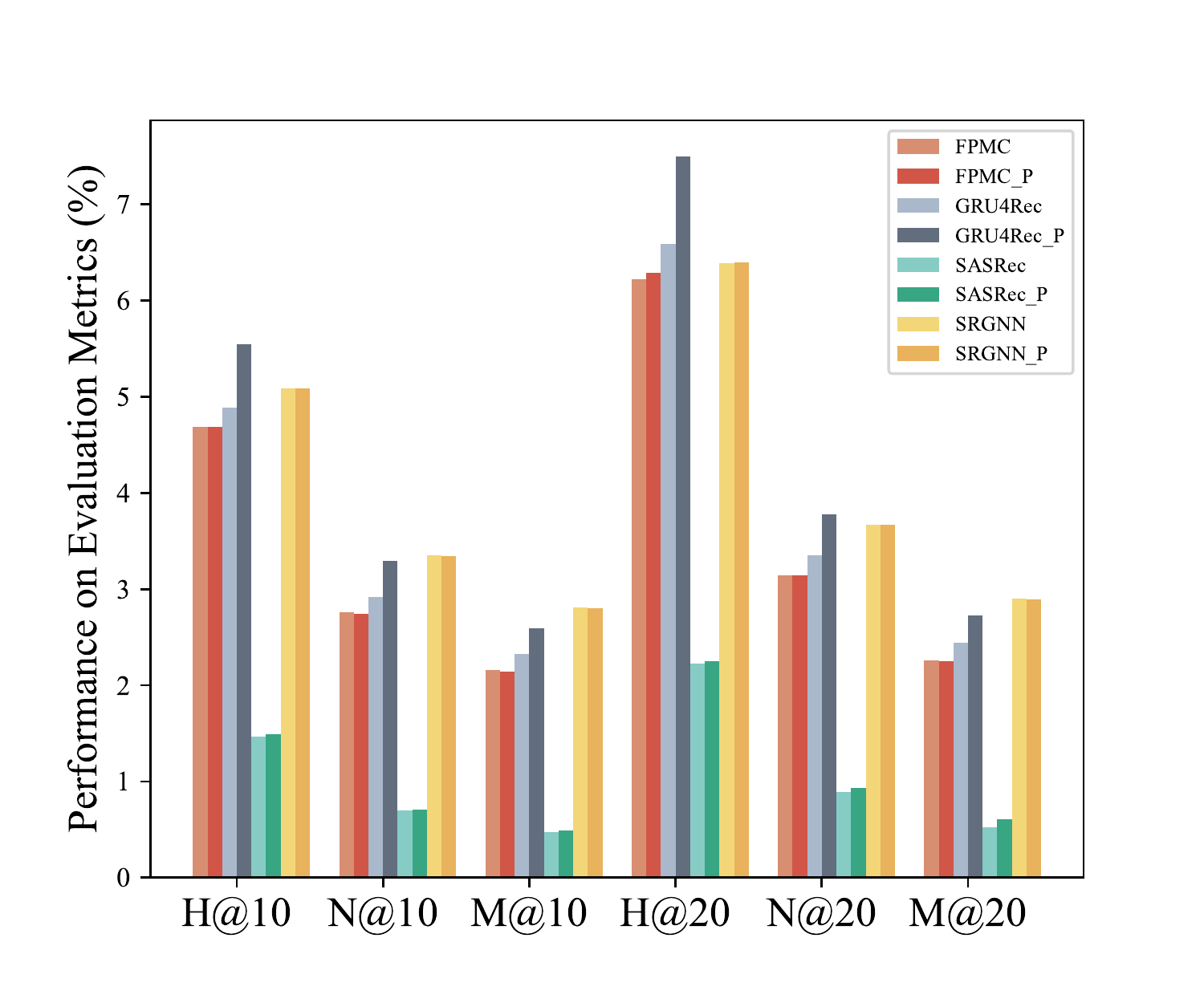}
\end{minipage}%
}%
\subfigure[Results on Pet.]{
\begin{minipage}[t]{0.4\linewidth}
\centering
\includegraphics[width=2.3in]{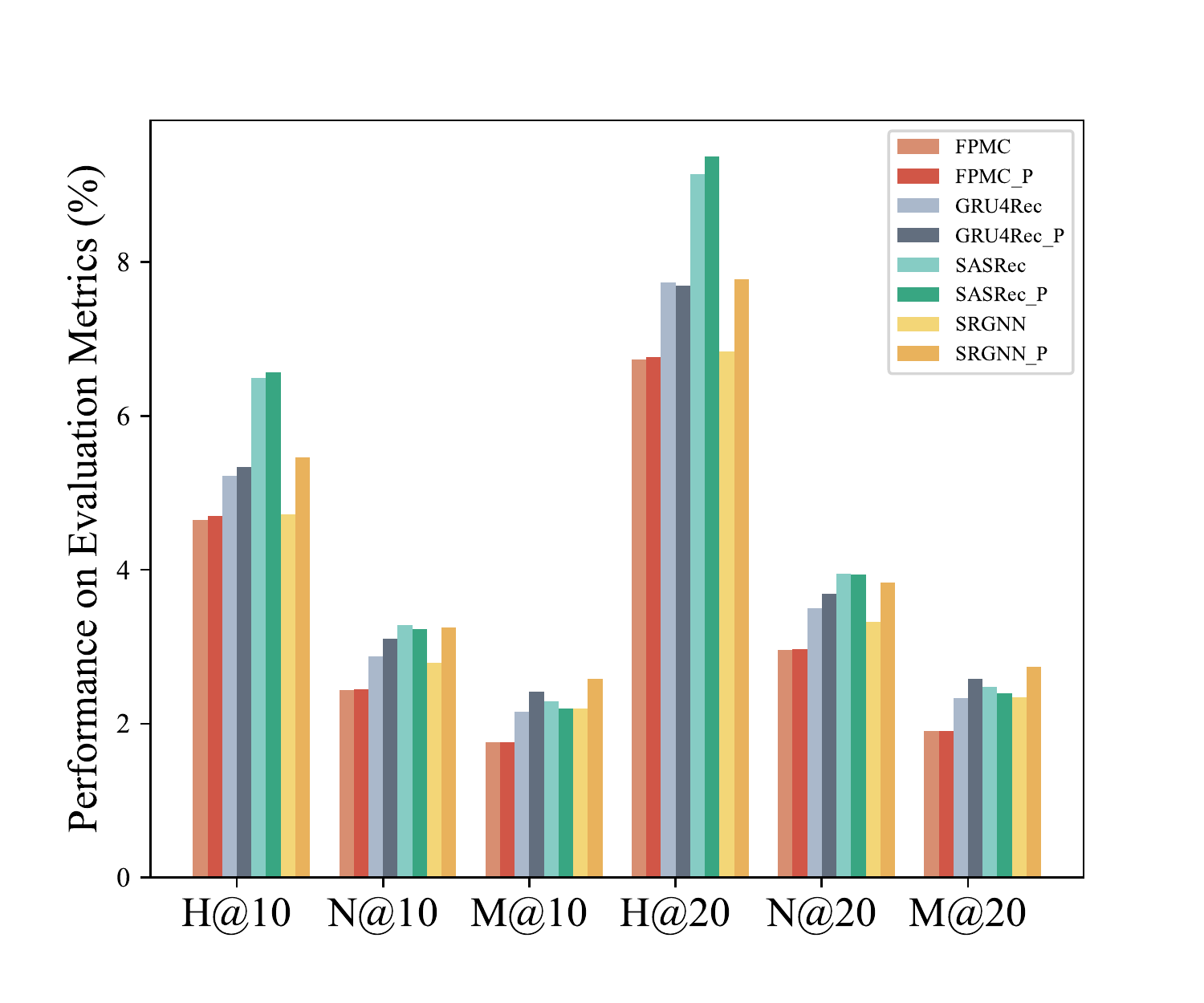}
\end{minipage}%
}%

\subfigure[Results on TH.]{
\begin{minipage}[t]{0.4\linewidth}
\centering
\includegraphics[width=2.3in]{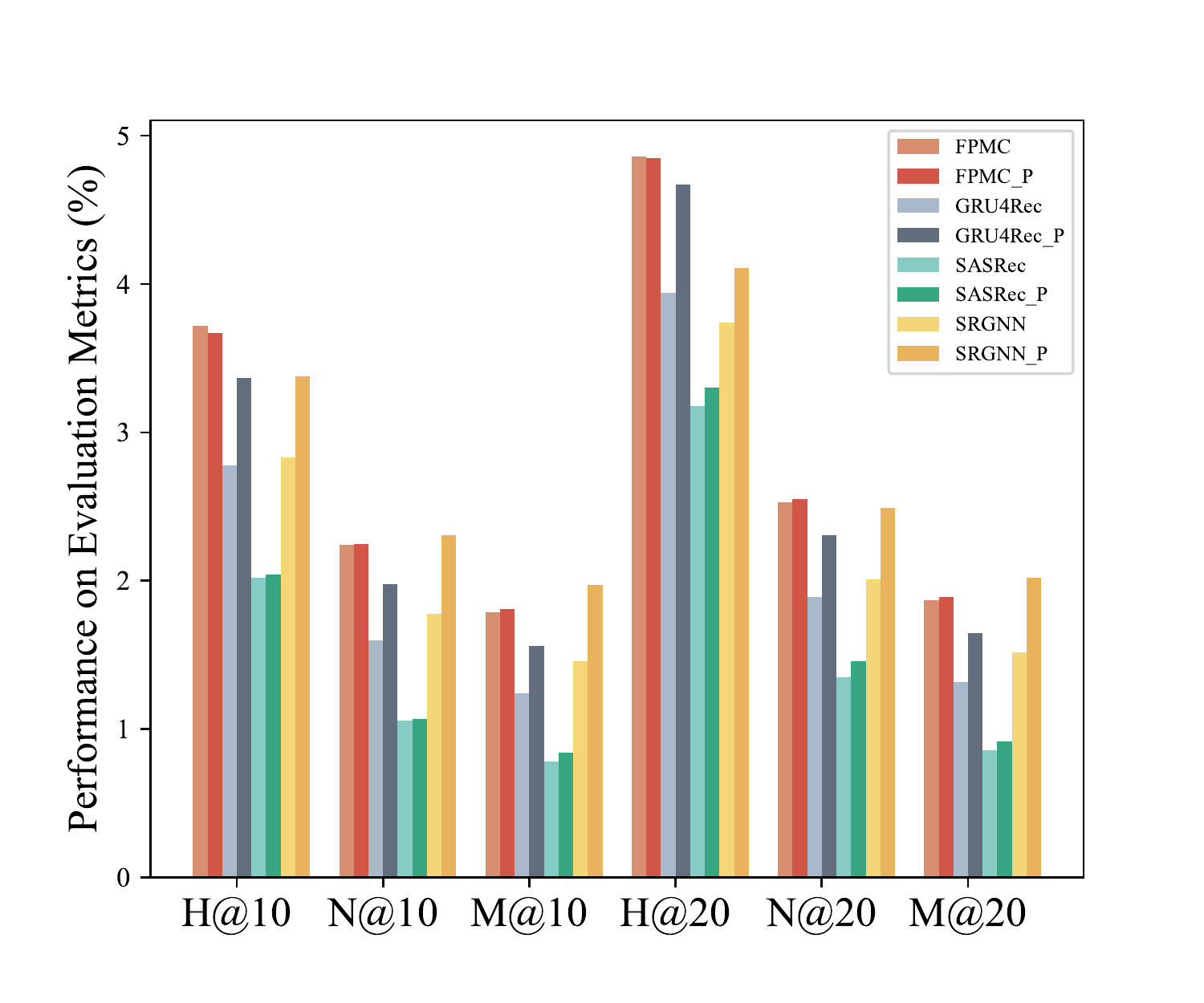}
\end{minipage}
}%
\subfigure[Results on MYbank.]{
\begin{minipage}[t]{0.4\linewidth}
\centering
\includegraphics[width=2.3in]{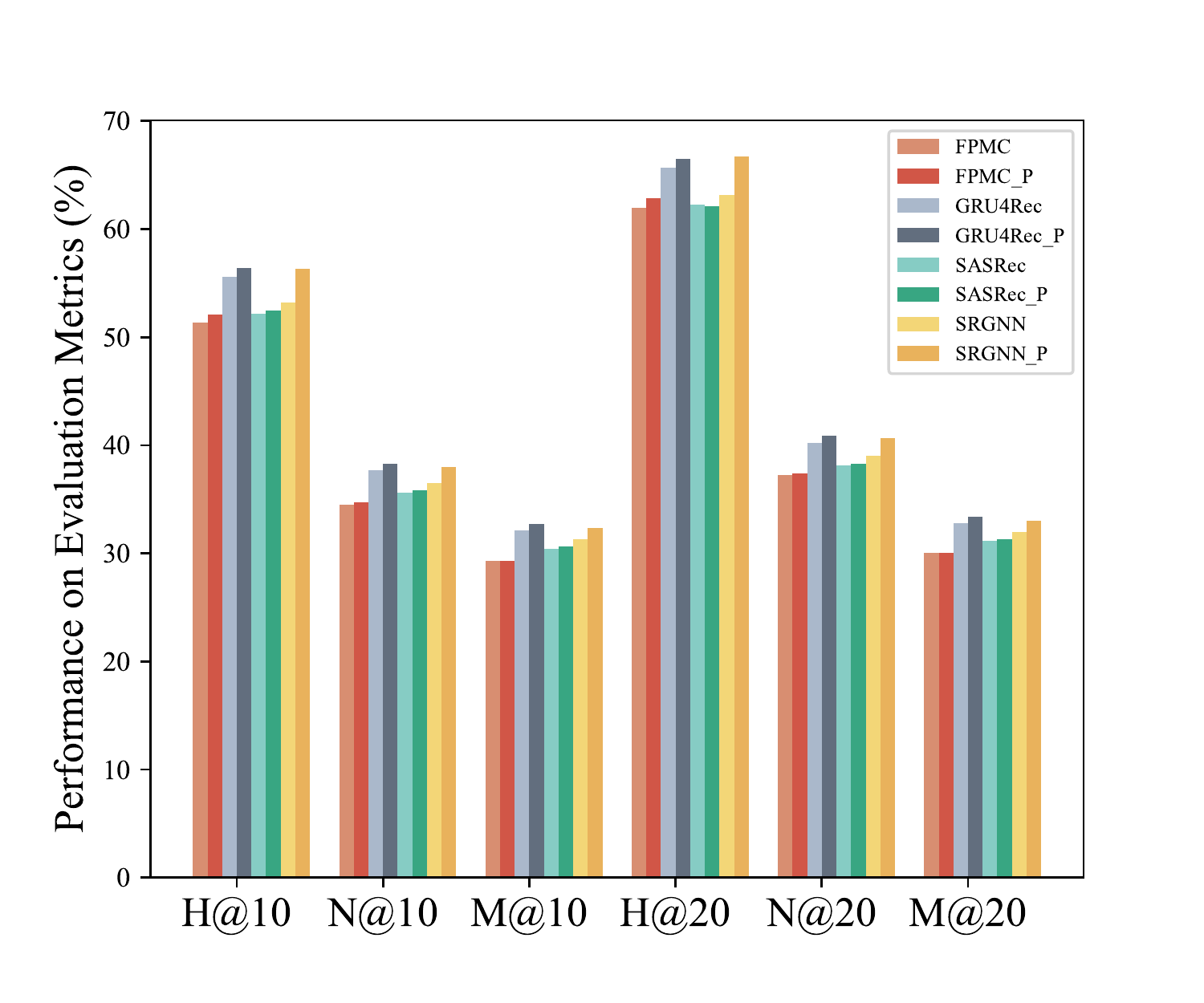}
\end{minipage}
}%
\centering
\caption{Results of the portability experiments on four datasets.}
\label{fig:portable_experiment}
\end{figure}

One of the main novelties of PHGR is that it is capable of capturing the items' hierarchical property on the space spanned by the ball. As discussed above, the Poincar\'{e} ball module is portable, which means is can be applied to most SR methods. To further demonstrate the effectiveness and portability of the Poincar\'{e} ball module, we replace the Euclidean space module within the FPMC, GRU4Rac, SARc, and SRGNN methods with the Poincar\'{e} ball module, i.e., the computations of all comparison methods are performed on the Poincar\'{e} ball. The results are presented in Figure \ref{fig:portable_experiment}, in which the methods with names followed by `-p' represents the remolded ones. 

\textbf{Analysis}: The results between the original methods and the remolded versions on different datasets are plotted in Figure \ref{fig:portable_experiment}. Firstly, we can conclude that the Poincar\'{e} ball module is portable to various SR methods according to our experiments. Furthermore, we can discover the Poincar\'{e} ball module works well with these SR models, not only with proposed PHGR. As observed, the remolded methods outperform their original versions with average improvements of 3.84\%, 4.72\%, 10.74\%, and 1.73\% on Beauty, Pet, T\&H, and MYbank, respectively, which demonstrates the effectiveness of the metric operation under Poincar\'{e} ball space. 

\subsection{Parameter Sensitive}
\begin{figure}[ht] 
\centering 
\includegraphics[width=0.65\linewidth]{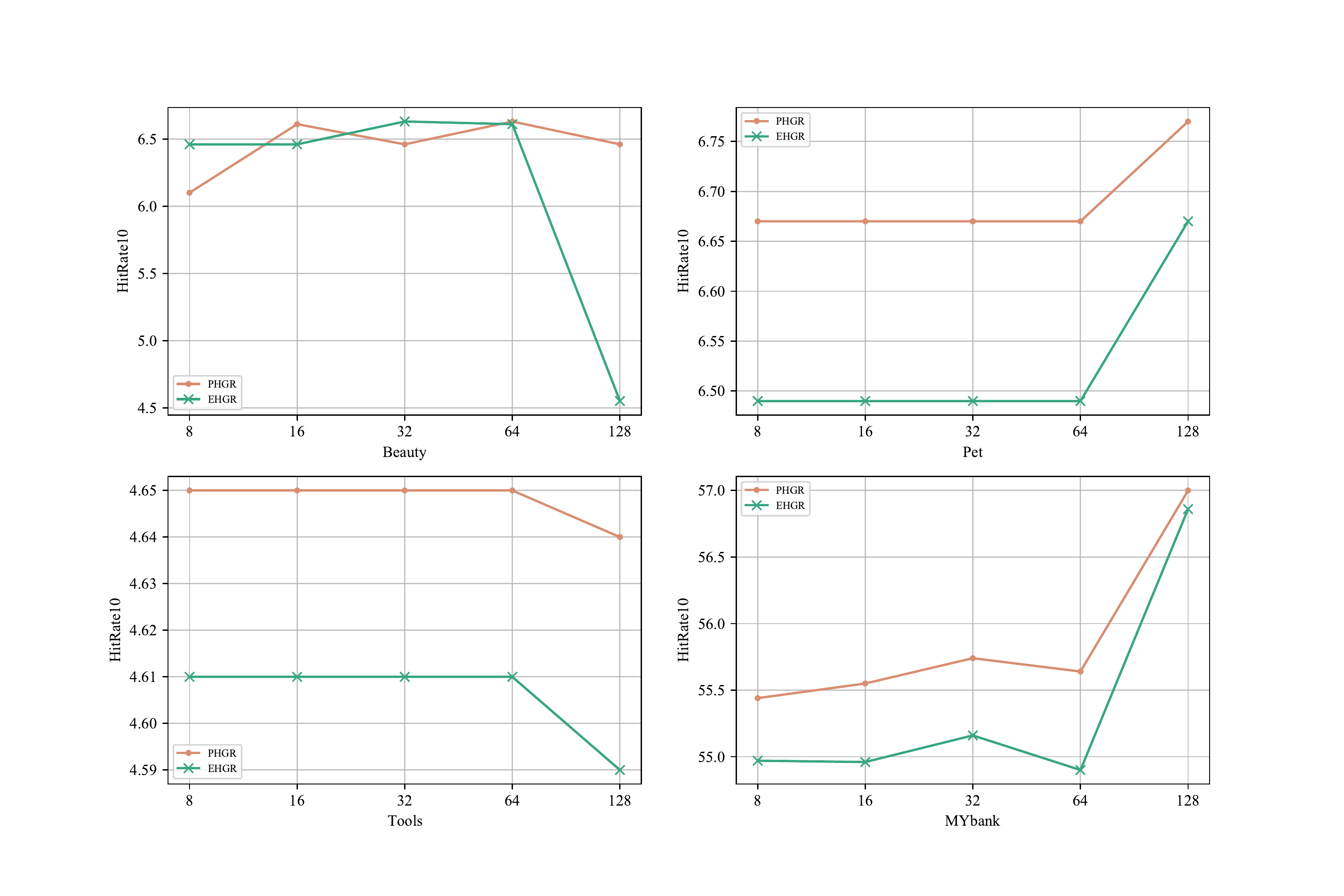}
\caption{The performance of PHGR and EHGR with different embedding size on four datasets.} 
\label{emb} 
\end{figure}

\begin{figure}[h] 
\centering 
\includegraphics[width=0.65\linewidth]{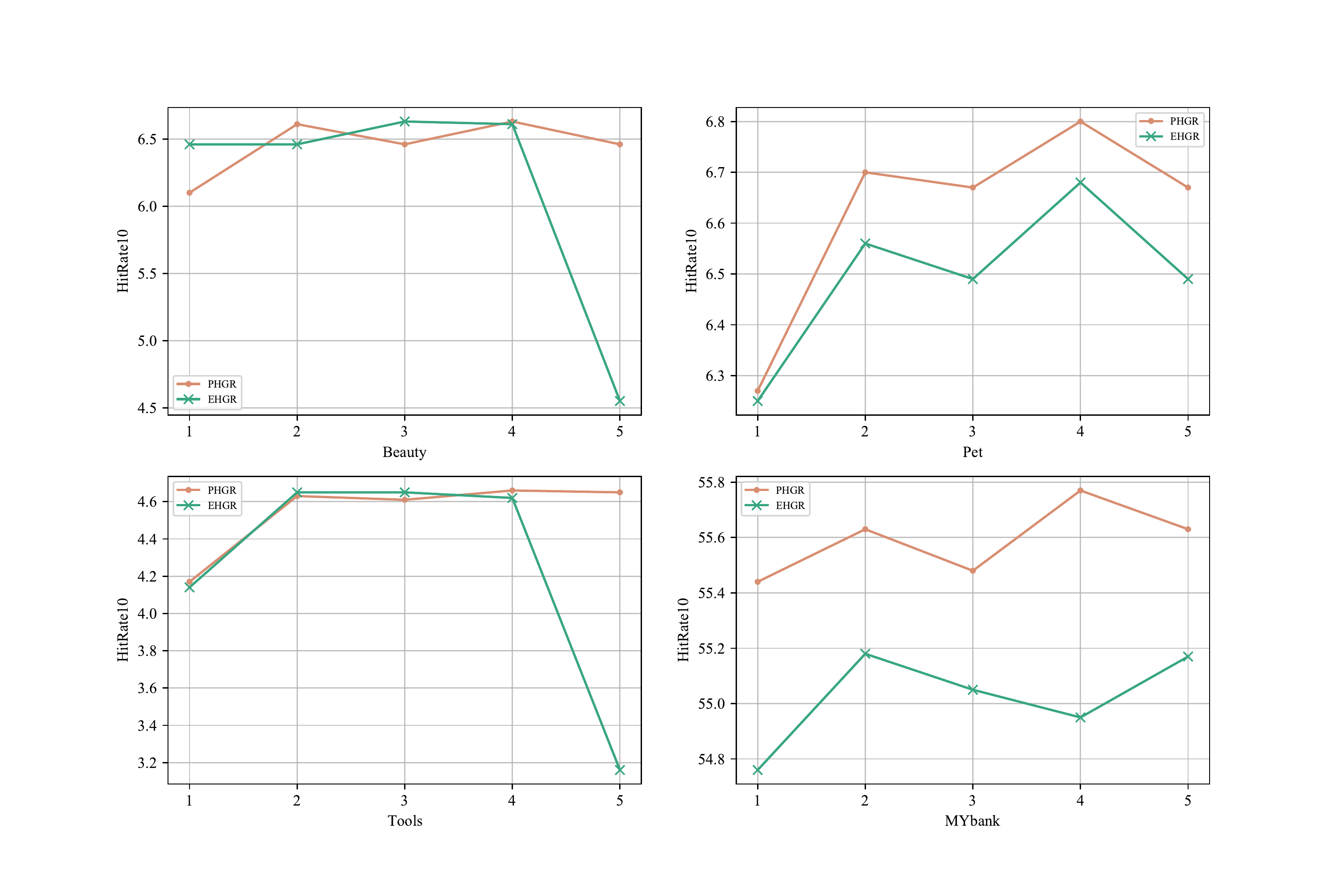}
\caption{The performance of PHGR and EHGR with different aggregation layer on four datasets.} 
\label{step} 
\end{figure}

In order to further investigate the power of the Poincar\'{e} ball module and provide a more comprehensive understanding of proposed PHGR in parameter sensitivity, we evaluate the performances of PHGR and EHGR under different choices of three important parameters. The evaluated parameters are: (1) the user and item embedding size $d$, which is essential as discussed in most sequential recommendation literature; (2) the number of the layers $L$ in the graph neural networks, which will result in over-smoothing or insufficient expressivity with improper values; and (3) the magnitude of the auxiliary loss $\omega$, which controls the contribution from the auxiliary learning to the ranking performance. The results are presented in Figures \ref{emb}, \ref{step} and \ref{loss}.

\begin{table}[H]
\centering
\caption{{The mean performance and its standard error of HitRate@10 on TH and Pet datasets.}}

\label{mean_std_table}
\resizebox{0.6\linewidth}{!}{ 
\begin{tabular}{ccccccc}
\hline
Method                & Dataset                & Embedding size & 8       & 16      & 32      & 64      \\ \hline
\multirow{4}{*}{EHGR} & \multirow{2}{*}{TH} & mean           & 6.25  & 6.56  & 6.49  & 6.68  \\ \cline{3-7} 
                      &                        & std            & 0.196 & 0.194 & 0.190 & 0.172 \\ \cline{2-7} 
                      & \multirow{2}{*}{Pet}   & mean           & 4.14  & 4.64  & 4.655 & 4.60   \\ \cline{3-7} 
                      &                        & std            & 0.135 & 0.128 & 0.121 & 0.10  \\ \hline
\multirow{4}{*}{PHGR} & \multirow{2}{*}{TH} & mean           & 6.27    & 6.70   & 6.67  & 6.80   \\ \cline{3-7} 
                      &                        & std            & 0.185 & 0.180 & 0.157 & 0.155 \\ \cline{2-7} 
                      & \multirow{2}{*}{Pet}   & mean           & 4.17  & 4.63  & 4.61  & 4.66  \\ \cline{3-7} 
                      &                        & std            & 0.109 & 0.105 & 0.097 & 0.086 \\ \hline
\end{tabular}
}
\end{table}

\textbf{Analysis:} 
Figure \ref{emb} depicts the performance of PHGR and EHGR with respect to a set of the embedding size $d$ in \{$8$, $16$, $32$, $64$, $128$\}. We can make two conclusions from the plot. First, PHGR outperforms EHGR under all values of $d$, which again demonstrates PHGR's superiority in SR scenarios. Second, the performance of both methods on all four datasets is insensitive to the settings of $d$ when $d$ is not larger than $64$, but the performance behaves differently on various datasets when $d$ increases to $128$. In particular, the performance of both methods increases on the denser MYBank and Pet datasets, whereas decreases sharply on the sparser Beauty and T\&H datasets. Such differences indicate that a larger $d$ is required for more expressive representation learning when the dataset is much informative and vise versa. 
During the experiments, PHGR and EHGR in TH and Pet did not change significantly with the varying of embedding sizes when the embedding size is not larger than 64, so we selected 100 different random seeds on these two datasets and recorded the mean and std values of the results over sets of experiments.
The mean performance and the corresponding std of EHGR and PHGR methods upon embedding size from 8 to 64 are reported in Table \ref{mean_std_table}.
As we can observe, the mean performance of HitRate@10 from PHGR method is superior to that of EHGR for each embedding size on both TH and Pet datasets. In addition, the standard error of the performance from PHGR is lower than EHGR, which indicates its relatively stability with the increase of the embedding size and implicitly suggests the advantage from the novel inner product operation.


\begin{figure}[] 
\centering 
\includegraphics[width=0.65\linewidth]{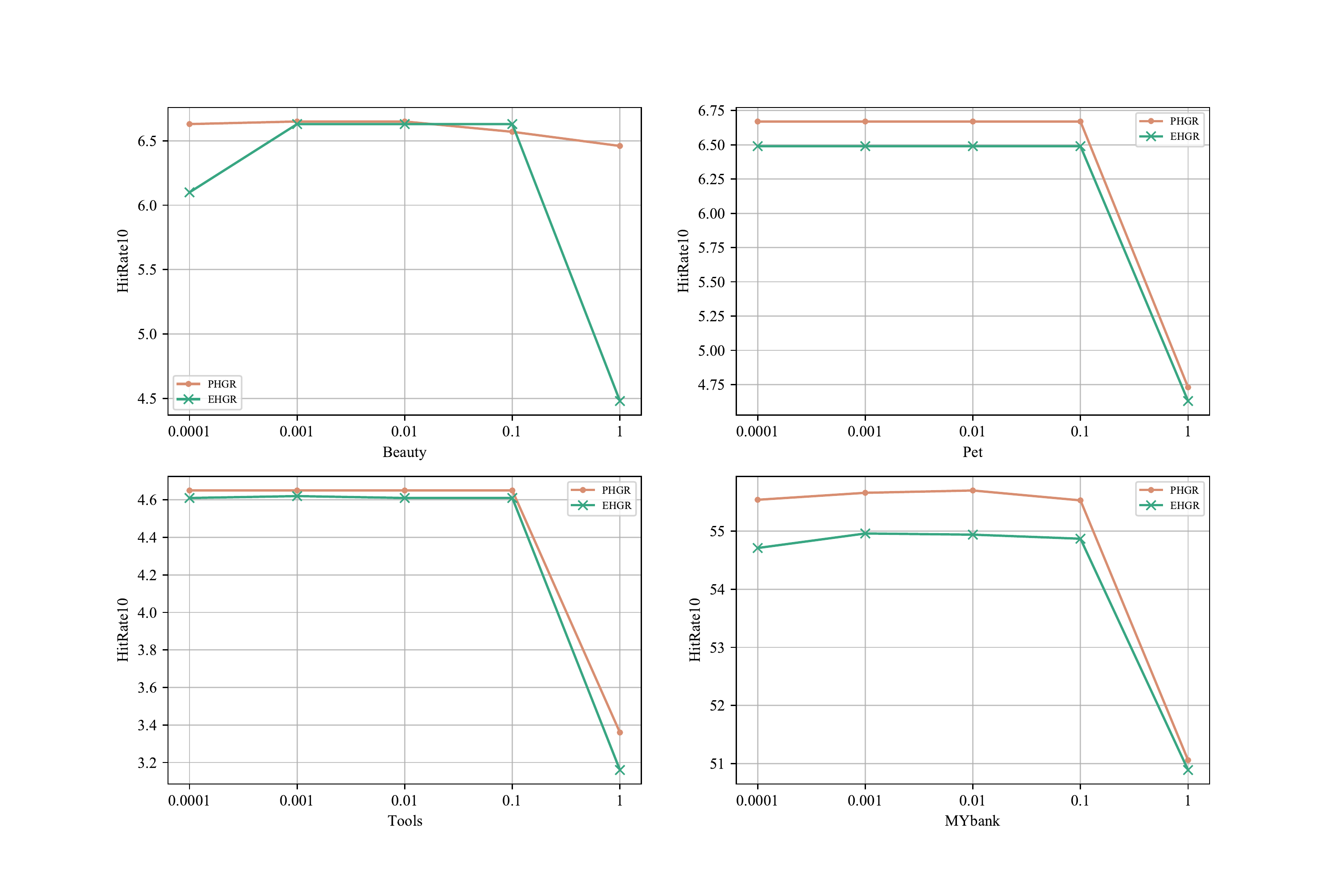}
\caption{The performance of PHGR and EHGR with different magnitude of the auxiliary loss on four datasets.} 
\label{loss} 
\end{figure}

In Figure \ref{step}, we report the performance of PHGR and EHGR with respective to a set of common settings \{$1$,$2$, $3$, $4$, $5$\} of the layers $L$ in the graph neural network. On average, PHGR outperforms EHGR, and the performance fluctuation of PHGR under varying $L$ values is more moderate than that of EHGR, which demonstrates the effectiveness and robustness of PHGR. Moreover, the observation that the performance of PHGR is relatively stable with varying $L$ indicates the powerful ability of PHGR in addressing the over-smoothing issue, which is a notorious problem in GNNs.

In Figure \ref{loss}, we analyze the performance of PHGR and EHGR with a set of representative values \{$0.0001$, $0.001$, $0.01$, $0.1$, $1$\} of $\omega$  in order to investigate the influence of the auxiliary loss, which contributes to the ranking outcome. 
On all datasets, it can be observed that small values of $\omega$ consistently lead to a stable and and better performance. However, as the  $\omega$ value increases, results on both model significantly decline probably due to the severe gradient conflict between two tasks. Therefore, in order to boost the best performance, it is important to make a trade-off between the hit ratio and ranking tasks when choosing the value of $\omega$.

\subsection{Representation Analysis}
The learning ability of the hierarchical structure in the data will affect the performance of the model, and the distance between the representation and the origin can reflect this type of structure. 
PHGR and EHGR are calculated in two different geometries, we using gyrovector space distance and tangent distance to calculate the distance from the target point to the origin.

\begin{figure}[H] 
\centering 
\includegraphics[width=0.6\linewidth]{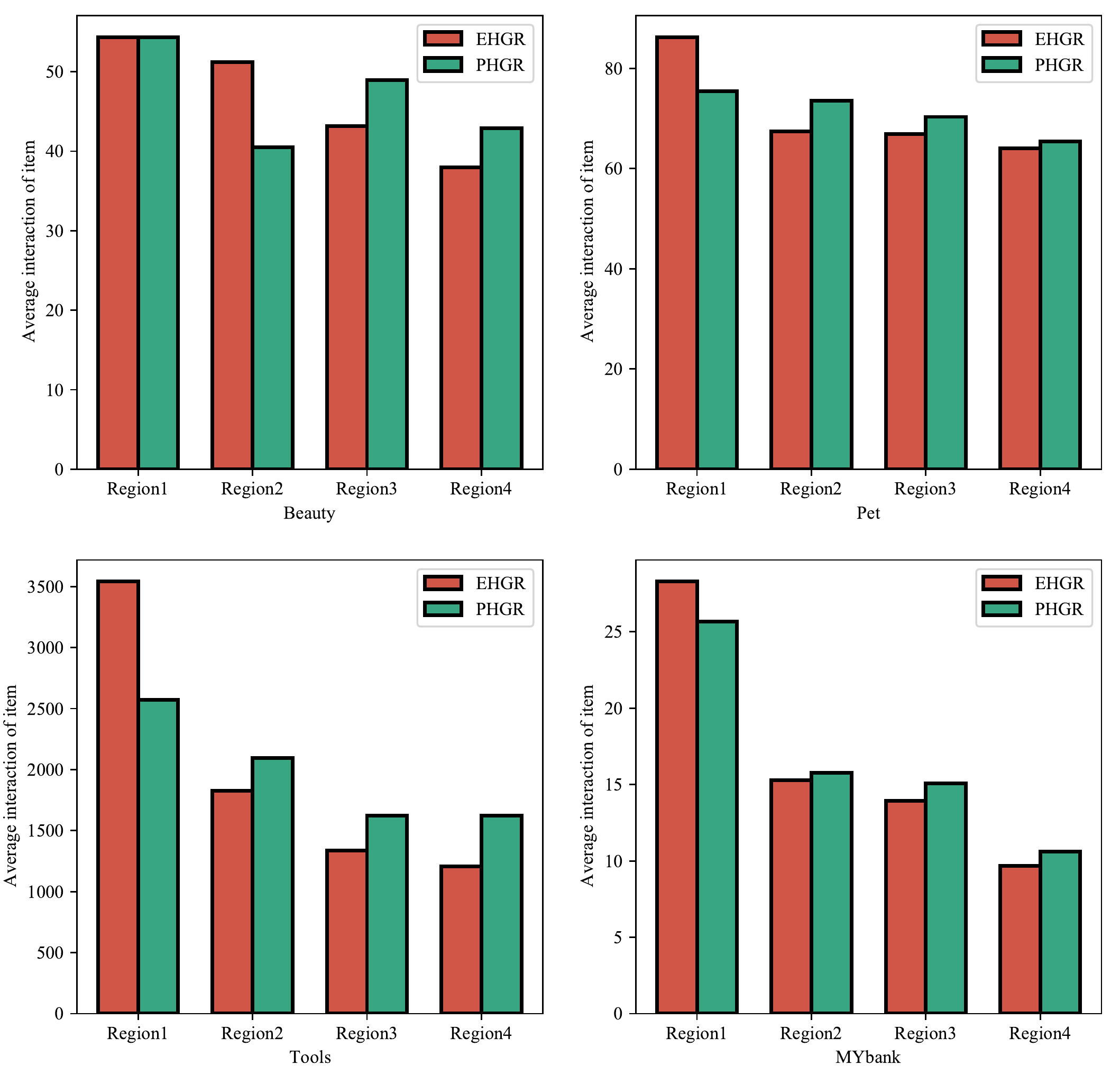}
\caption{Hierarchical representation analysis on four datasets.} 
\label{region} 
\end{figure}

First, we set up three boundaries in Euclidean space and Poincar\'{e} space, and divide the representation into four regions according to their distance from the origin.
For example, the item in region 1 is the closest to the origin, whereas the item in region 4 is the farthest from the origin. 
To intuitively reflect the different popularity of items in different regions, we count the interaction times of nodes in all regions on the four datasets. Then, we visualize the statistics in Figure \ref{region}.

As can be observed in Figure \ref{region}, the average number of interactions between items from region 1 to region 4 decreases, which suggests both approaches, EHGR and PHGR, are capable of modeling the hierarchical structure of sequence behavior.

Furthermore, in all datasets, the average number of item's interactions with PHGR in region 1 is higher than that with EHGR, while the average number of interactions with EHGR in region 3 and 4 is higher than that with PHGR.
PHGR distinguishes items of different popularity better than EHGR, emphasizing that hyperbolic space is more suitable for embedding hierarchical data than Euclidean space.

\subsection{Case Study of Poincar\'{e} Attention Illustration}
\begin{figure}[H] 
\centering 
\includegraphics[width=1.0\linewidth]{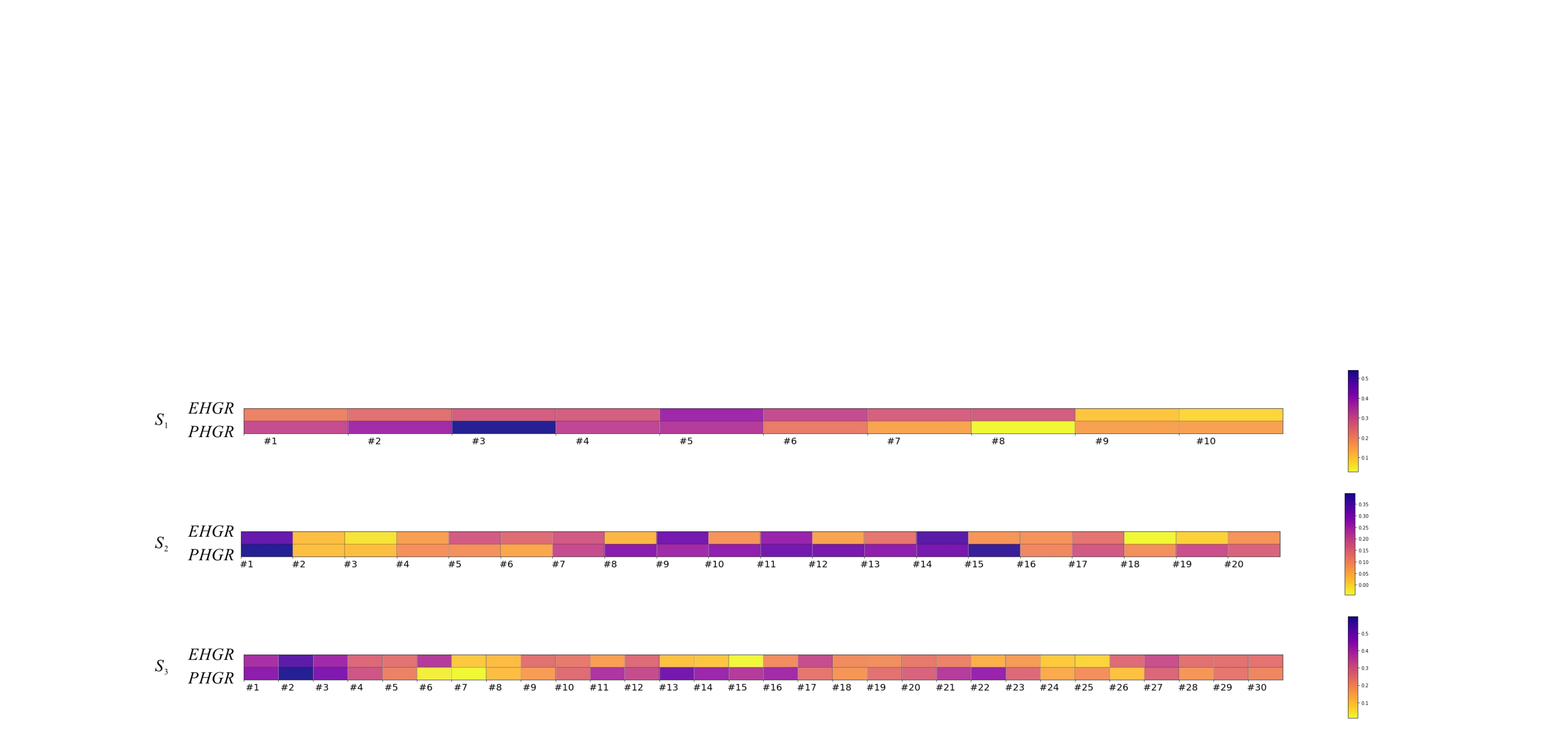}
\caption{The illustration of computed attention weights under euclidean and hyperbolic spaces.} 
\label{att_map} 
\end{figure}

To further understanding the Poincar\'{e} attention mechanism, we visualize the attention weight between user behaviors as shown in Figure \ref{att_map}, which reflects the different item influences within the same sequence on the two models(EHGR and PHGR). We randomly select three different behavior sequences of length 10($S_1$), 20($S_2$) and 30($S_3$) respectively from MYbank dataset(test set). For the heatmaps of sequences in the same group $S_i$ , the above one is the attention weight between the related items and the next item user most likely to click predicted by EHGR, while the following one is the corresponding attention weight modeled by PHGR. According to the heatmap, user behaviors in the same sequence contribute differently to the recommendation result.
In comparison to EHGR, PHGR's attention weight is more distinguishable and obviously higher in many key behaviors.
Particularly, PHGR will give higher scores with the increase of the scores provided by EHGR, which indicates that PHGR is more distinguishable on item importance. The case study result implies that hyperbolic space can effectively represent the hierarchical structure of data, the influence of each behavior among sequence is better measured.

\subsection{Training Efficiency}
In order to provide a detail efficiency evaluation, we analyze the time complexity according to the main paper section Experimental Settings \ref{Experimental Settings}. All methods are tested on the same machine, which consists of a single NVIDIA Tesla V100 GPU with 32GB memory and Intel(R) Xeon(R) Platinum 8163 CPU @2.50GHz with 128G RAM. The performance of graph-based models  (SRGNN, GC-SAN, LESSR)  greatly improves in terms of adopting our proposed novel attention mechanism based on Poincare graph neural networks instead of involving GRU modules and without required additional computational costs. The number of parameters for SRGNN, GC-SAN, LESSR and PHGR are similar, corresponding to 8.72M, 2.80M, 8.97M and 8.78M respectively. Specifically, although GC-SAN has less parameters but it requires more training time. As we record, the training efficiency is 438.5min, 556min, 451min and 461 min for SRGNN, GC-SAN, LESSR and PHGR correspondingly. PHGR is with a average size and computational costs compared to most existed methods.

\section{Conclusions}
\label{conclusion}
{In this paper, to model the sequential dependency and hierarchy-like structure contained in the data of SR scenarios simultaneously, we propose a Poincar\'{e}-based graph neural network named PHGR. By defining a novel hyperbolic inner product operator, the global heterogeneous graph representation learning and local homogeneous graph representation learning are conducted in Poincar\'{e} ball space to capture the hierarchical information. Besides, two sequential attention operations in , i.e., long-view temporal attention and short-view temporal attention are explicitly introduced to capture the sequential dependency information. Empirical experiments on three widely-used and one real-world industrial datasets demonstrate the efficiency of PHGR and the extensibility of the proposed Poincar\'{e} operation module for other SOTA approaches in SR scenarios. }

\bibliographystyle{unsrt}
\bibliography{sample-base}

\begin{thebibliography}{10}

\bibitem{rendle2010factorizing}
Steffen Rendle, Christoph Freudenthaler, and Lars Schmidt-Thieme.
\newblock Factorizing personalized markov chains for next-basket
  recommendation.
\newblock In {\em Proceedings of the 19th international conference on World
  wide web}, pages 811--820, 2010.

\bibitem{wang2015learning}
Pengfei Wang, Jiafeng Guo, Yanyan Lan, Jun Xu, Shengxian Wan, and Xueqi Cheng.
\newblock Learning hierarchical representation model for nextbasket
  recommendation.
\newblock In {\em Proceedings of the 38th International ACM SIGIR conference on
  Research and Development in Information Retrieval}, pages 403--412, 2015.

\bibitem{he2016fusing}
Ruining He and Julian McAuley.
\newblock Fusing similarity models with markov chains for sparse sequential
  recommendation.
\newblock In {\em 2016 IEEE 16th International Conference on Data Mining
  (ICDM)}, pages 191--200. IEEE, 2016.

\bibitem{hidasi2015session}
Bal{\'a}zs Hidasi, Alexandros Karatzoglou, Linas Baltrunas, and Domonkos Tikk.
\newblock Session-based recommendations with recurrent neural networks.
\newblock {\em arXiv preprint arXiv:1511.06939}, 2015.

\bibitem{li2017neural}
Jing Li, Pengjie Ren, Zhumin Chen, Zhaochun Ren, Tao Lian, and Jun Ma.
\newblock Neural attentive session-based recommendation.
\newblock In {\em Proceedings of the 2017 ACM on Conference on Information and
  Knowledge Management}, pages 1419--1428, 2017.

\bibitem{cui2018mv}
Qiang Cui, Shu Wu, Qiang Liu, Wen Zhong, and Liang Wang.
\newblock {MV-RNN}: A multi-view recurrent neural network for sequential
  recommendation.
\newblock {\em IEEE Transactions on Knowledge and Data Engineering},
  32(2):317--331, 2018.

\bibitem{kang2018self}
Wang-Cheng Kang and Julian McAuley.
\newblock Self-attentive sequential recommendation.
\newblock In {\em 2018 IEEE International Conference on Data Mining (ICDM)},
  pages 197--206. IEEE, 2018.

\bibitem{zheng2020sentiment}
Lin Zheng, Naicheng Guo, Weihao Chen, Jin Yu, and Dazhi Jiang.
\newblock Sentiment-guided sequential recommendation.
\newblock In {\em Proceedings of the 43rd International ACM SIGIR Conference on
  Research and Development in Information Retrieval}, pages 1957--1960, 2020.

\bibitem{fan2021lighter}
Xinyan Fan, Zheng Liu, Jianxun Lian, Wayne~Xin Zhao, Xing Xie, and Ji-Rong Wen.
\newblock Lighter and better: Low-rank decomposed self-attention networks for
  next-item recommendation.
\newblock In {\em Proceedings of the 44th International ACM SIGIR Conference on
  Research and Development in Information Retrieval}, pages 1733--1737, 2021.

\bibitem{wu2019session}
Shu Wu, Yuyuan Tang, Yanqiao Zhu, Liang Wang, Xing Xie, and Tieniu Tan.
\newblock Session-based recommendation with graph neural networks.
\newblock In {\em Proceedings of the AAAI Conference on Artificial
  Intelligence}, volume~33, pages 346--353, 2019.

\bibitem{wu2019personalizing}
Shu Wu, Mengqi Zhang, Xin Jiang, Xu~Ke, and Liang Wang.
\newblock Personalizing graph neural networks with attention mechanism for
  session-based recommendation.
\newblock {\em arXiv preprint arXiv:1910.08887}, 2019.

\bibitem{xu2019graph}
Chengfeng Xu, Pengpeng Zhao, Yanchi Liu, Victor~S Sheng, Jiajie Xu, Fuzhen
  Zhuang, Junhua Fang, and Xiaofang Zhou.
\newblock Graph contextualized self-attention network for session-based
  recommendation.
\newblock In {\em IJCAI}, volume~19, pages 3940--3946, 2019.

\bibitem{ravasz2003hierarchical}
Erzs{\'e}bet Ravasz and Albert-L{\'a}szl{\'o} Barab{\'a}si.
\newblock Hierarchical organization in complex networks.
\newblock {\em Physical review E}, 67(2):026112, 2003.

\bibitem{wang2018exploring}
Suhang Wang, Jiliang Tang, Yilin Wang, and Huan Liu.
\newblock Exploring hierarchical structures for recommender systems.
\newblock {\em IEEE Transactions on Knowledge and Data Engineering},
  30(6):1022--1035, 2018.

\bibitem{ma2019hierarchical}
Chen Ma, Peng Kang, and Xue Liu.
\newblock Hierarchical gating networks for sequential recommendation.
\newblock In {\em Proceedings of the 25th ACM SIGKDD international conference
  on knowledge discovery \& data mining}, pages 825--833, 2019.

\bibitem{li2020hierarchical}
Xingchen Li, Xiang Wang, Xiangnan He, Long Chen, Jun Xiao, and Tat-Seng Chua.
\newblock Hierarchical fashion graph network for personalized outfit
  recommendation.
\newblock In {\em Proceedings of the 43rd International ACM SIGIR Conference on
  Research and Development in Information Retrieval}, pages 159--168, 2020.

\bibitem{bronstein2017geometric}
Michael~M Bronstein, Joan Bruna, Yann LeCun, Arthur Szlam, and Pierre
  Vandergheynst.
\newblock Geometric deep learning: Going beyond euclidean data.
\newblock {\em IEEE Signal Processing Magazine}, 34(4):18--42, 2017.

\bibitem{sala2018representation}
Frederic Sala, Chris De~Sa, Albert Gu, and Christopher R{\'e}.
\newblock Representation tradeoffs for hyperbolic embeddings.
\newblock In {\em International conference on machine learning}, pages
  4460--4469. PMLR, 2018.

\bibitem{gulcehre2018hyperbolic}
Caglar Gulcehre, Misha Denil, Mateusz Malinowski, Ali Razavi, Razvan Pascanu,
  Karl~Moritz Hermann, Peter Battaglia, Victor Bapst, David Raposo, Adam
  Santoro, et~al.
\newblock Hyperbolic attention networks.
\newblock {\em arXiv preprint arXiv:1805.09786}, 2018.

\bibitem{krioukov2008efficient}
Dmitri Krioukov, Fragkiskos Papadopoulos, Mari{\'a}n Bogun{\'a}, Amin Vahdat,
  et~al.
\newblock Efficient navigation in scale-free networks embedded in hyperbolic
  metric spaces.
\newblock Technical report, arXiv cond-mat. stat-mech/0805.1266, 2008.

\bibitem{krioukov2010hyperbolic}
Dmitri Krioukov, Fragkiskos Papadopoulos, Maksim Kitsak, Amin Vahdat, and
  Mari{\'a}n Bogun{\'a}.
\newblock Hyperbolic geometry of complex networks.
\newblock {\em Physical Review E}, 82(3):036106, 2010.

\bibitem{chamberlain2019scalable}
Benjamin~Paul Chamberlain, Stephen~R Hardwick, David~R Wardrope, Fabon Dzogang,
  Fabio Daolio, and Sa{\'u}l Vargas.
\newblock Scalable hyperbolic recommender systems.
\newblock {\em arXiv preprint arXiv:1902.08648}, 2019.

\bibitem{Ky2020}
Kyuyong Shin, Young{-}Jin Park, Kyung{-}Min Kim, and Sunyoung Kwon.
\newblock Multi-manifold learning for large-scale targeted advertising system.
\newblock {\em arXiv preprint arXiv: 2007.02334}, 2020.

\bibitem{ch2021}
Chen Ma, Liheng Ma, Yingxue Zhang, Haolun Wu, Xue Liu, and Mark Coates.
\newblock Knowledge-enhanced top-k recommendation in poincar{\'{e}} ball.
\newblock In {\em Thirty-Fifth {AAAI} Conference on Artificial Intelligence,
  {AAAI} 2021, Thirty-Third Conference on Innovative Applications of Artificial
  Intelligence, {IAAI} 2021, The Eleventh Symposium on Educational Advances in
  Artificial Intelligence, {EAAI} 2021, Virtual Event, February 2-9, 2021},
  pages 4285--4293, 2021.

\bibitem{li2021hsr}
Anchen Li and Bo~Yang.
\newblock {HSR:} hyperbolic social recommender.
\newblock {\em arXiv preprint arXiv: 2102.09389}, abs/2102.09389, 2021.

\bibitem{lin2020fissa}
Jing Lin, Weike Pan, and Zhong Ming.
\newblock {FISSA}: Fusing item similarity models with self-attention networks
  for sequential recommendation.
\newblock In {\em Fourteenth ACM Conference on Recommender Systems}, pages
  130--139, 2020.

\bibitem{wang2020knowledge}
Baocheng Wang and Wentao Cai.
\newblock Knowledge-enhanced graph neural networks for sequential
  recommendation.
\newblock {\em Information}, 11(8):388, 2020.

\bibitem{wang2020global}
Ziyang Wang, Wei Wei, Gao Cong, Xiao-Li Li, Xian-Ling Mao, and Minghui Qiu.
\newblock Global context enhanced graph neural networks for session-based
  recommendation.
\newblock In {\em Proceedings of the 43rd International ACM SIGIR Conference on
  Research and Development in Information Retrieval}, pages 169--178, 2020.

\bibitem{yi2019sampling}
Xinyang Yi, Ji~Yang, Lichan Hong, Derek~Zhiyuan Cheng, Lukasz Heldt, Aditee
  Kumthekar, Zhe Zhao, Li~Wei, and Ed~Chi.
\newblock Sampling-bias-corrected neural modeling for large corpus item
  recommendations.
\newblock In {\em Proceedings of the 13th ACM Conference on Recommender
  Systems}, pages 269--277, 2019.

\bibitem{balazevic2019multi}
Ivana Balazevic, Carl Allen, and Timothy Hospedales.
\newblock Multi-relational poincar{\'e} graph embeddings.
\newblock {\em Advances in Neural Information Processing Systems}, 32, 2019.

\bibitem{ma2020memory}
Chen Ma, Liheng Ma, Yingxue Zhang, Jianing Sun, Xue Liu, and Mark Coates.
\newblock Memory augmented graph neural networks for sequential recommendation.
\newblock In {\em Proceedings of the AAAI Conference on Artificial
  Intelligence}, volume~34, pages 5045--5052, 2020.

\bibitem{chen2020handling}
Tianwen Chen and Raymond Chi-Wing Wong.
\newblock Handling information loss of graph neural networks for session-based
  recommendation.
\newblock In {\em Proceedings of the 26th ACM SIGKDD International Conference
  on Knowledge Discovery \& Data Mining}, pages 1172--1180, 2020.

\bibitem{duin2010non}
Robert~PW Duin and El{\.z}bieta P{\k{e}}kalska.
\newblock Non-euclidean dissimilarities: Causes and informativeness.
\newblock In {\em Joint IAPR International Workshops on Statistical Techniques
  in Pattern Recognition (SPR) and Structural and Syntactic Pattern Recognition
  (SSPR)}, pages 324--333. Springer, 2010.

\bibitem{tifrea2018poincar}
Alexandru Tifrea, Gary B{\'e}cigneul, and Octavian-Eugen Ganea.
\newblock Poincar{\'e} glove: Hyperbolic word embeddings.
\newblock {\em arXiv preprint arXiv:1810.06546}, 2018.

\bibitem{nickel2017poincare}
Maximillian Nickel and Douwe Kiela.
\newblock Poincar{\'e} embeddings for learning hierarchical representations.
\newblock {\em Advances in neural information processing systems},
  30:6338--6347, 2017.

\bibitem{ganea2018hyperbolic}
Octavian-Eugen Ganea, Gary B{\'e}cigneul, and Thomas Hofmann.
\newblock Hyperbolic neural networks.
\newblock {\em arXiv preprint arXiv:1805.09112}, 2018.

\bibitem{chami2019hyperbolic}
Ines Chami, Zhitao Ying, Christopher R{\'e}, and Jure Leskovec.
\newblock Hyperbolic graph convolutional neural networks.
\newblock {\em Advances in neural information processing systems},
  32:4868--4879, 2019.

\bibitem{mathieu2019continuous}
Emile Mathieu, Charline~Le Lan, Chris~J Maddison, Ryota Tomioka, and Yee~Whye
  Teh.
\newblock Continuous hierarchical representations with poincar{\'e} variational
  auto-encoders.
\newblock {\em arXiv preprint arXiv:1901.06033}, 2019.

\bibitem{ovinnikov2019poincar}
Ivan Ovinnikov.
\newblock Poincar{\'e} wasserstein autoencoder.
\newblock {\em arXiv preprint arXiv:1901.01427}, 2019.

\bibitem{skopek2019mixed}
Ondrej Skopek, Octavian-Eugen Ganea, and Gary B{\'e}cigneul.
\newblock Mixed-curvature variational autoencoders.
\newblock {\em arXiv preprint arXiv:1911.08411}, 2019.

\bibitem{chami2020low}
Ines Chami, Adva Wolf, Da-Cheng Juan, Frederic Sala, Sujith Ravi, and
  Christopher R{\'e}.
\newblock Low-dimensional hyperbolic knowledge graph embeddings.
\newblock {\em arXiv preprint arXiv:2005.00545}, 2020.

\bibitem{choudhary2021self}
Nurendra Choudhary, Nikhil Rao, Sumeet Katariya, Karthik Subbian, and Chandan~K
  Reddy.
\newblock Self-supervised hyperboloid representations from logical queries over
  knowledge graphs.
\newblock In {\em Proceedings of the Web Conference 2021}, pages 1373--1384,
  2021.

\bibitem{feng2020hme}
Shanshan Feng, Lucas~Vinh Tran, Gao Cong, Lisi Chen, Jing Li, and Fan Li.
\newblock {HME}: A hyperbolic metric embedding approach for next-poi
  recommendation.
\newblock In {\em Proceedings of the 43rd International ACM SIGIR Conference on
  Research and Development in Information Retrieval}, pages 1429--1438, 2020.

\bibitem{mirvakhabova2020performance}
Leyla Mirvakhabova, Evgeny Frolov, Valentin Khrulkov, Ivan Oseledets, and
  Alexander Tuzhilin.
\newblock Performance of hyperbolic geometry models on top-n recommendation
  tasks.
\newblock In {\em Fourteenth ACM Conference on Recommender Systems}, pages
  527--532, 2020.

\bibitem{wang2021hypersorec}
Hao Wang, Defu Lian, Hanghang Tong, Qi~Liu, Zhenya Huang, and Ennhong Chen.
\newblock Hypersorec: Exploiting hyperbolic user and item representations with
  multiple aspects for social-aware recommendation.
\newblock 2021.

\bibitem{zhang2019heterogeneous}
Chuxu Zhang, Dongjin Song, Chao Huang, Ananthram Swami, and Nitesh~V Chawla.
\newblock Heterogeneous graph neural network.
\newblock In {\em Proceedings of the 25th ACM SIGKDD International Conference
  on Knowledge Discovery \& Data Mining}, pages 793--803, 2019.

\bibitem{chamberlain2017neural}
Benjamin~Paul Chamberlain, James Clough, and Marc~Peter Deisenroth.
\newblock Neural embeddings of graphs in hyperbolic space.
\newblock {\em arXiv preprint arXiv:1705.10359}, 2017.

\bibitem{vaswani2017attention}
Ashish Vaswani, Noam Shazeer, Niki Parmar, Jakob Uszkoreit, Llion Jones,
  Aidan~N Gomez, Lukasz Kaiser, and Illia Polosukhin.
\newblock Attention is all you need.
\newblock {\em arXiv preprint arXiv:1706.03762}, 2017.

\bibitem{guo2021hcgr}
Naicheng Guo, Xiaolei Liu, Shaoshuai Li, Qiongxu Ma, Yunan Zhao, Bing Han, Lin
  Zheng, Kaixin Gao, and Xiaobo Guo.
\newblock {HCGR}: Hyperbolic contrastive graph representation learning for
  session-based recommendation.
\newblock {\em arXiv preprint arXiv:2107.05366}, 2021.

\end{thebibliography}

\end{document}